\documentclass[11pt,onecolumn]{IEEEtran}

\usepackage{citesort}
\usepackage[fleqn]{amsmath}
\usepackage{ascmac}
\usepackage[dvips]{graphicx}
\usepackage{psfrag}
\usepackage{amsmath}
\usepackage{amsbsy}
\usepackage{amssymb}
\usepackage{latexsym}
\usepackage{subfigure}
\usepackage{color}
\usepackage{amsthm} 

 \newtheorem{proposition}{Proposition}
 \newtheorem{algorithm}{Algorithm}
 \newtheorem{definition}{Definition}
 \newtheorem{theorem}{Theorem}
 \newtheorem{lemma}{Lemma}
 \newtheorem{corollary}{Corollary}
 \newtheorem{fact}{Fact}
 \newtheorem{remark}{Remark}
 \newtheorem{observation}{Observation}
 \newtheorem{example}{Example}

\newcommand{\migip}{\hspace*{\fill} $\Box $}

\newcommand{\signal}[1]{{\boldsymbol{#1}}}

\newcommand{\norm}[1]{\left\|#1\right\|}
\newcommand{\abs}[1]{\left|#1\right|}

\newcommand{\real}{{\mathbb R}}

\newcommand{\refeq}[1]{(\ref{#1})}

\newcommand{\argmin}{\operatornamewithlimits{argmin}}
\newcommand{\argmax}{\operatornamewithlimits{argmax}}
\newcommand{\Rep}[1]{(\mathcal{Q}_{\varepsilon}^{#1})}
\newcommand{\Pc}[1]{(\mathcal{P}_c^{#1})}
\newcommand{\Ql}[1]{(\mathcal{L}_{\lambda}^{#1})}
\newcommand{\Pct}[1]{(\widetilde{\mathcal{P}}_c^{#1})}
\newcommand{\Qlt}[1]{(\widetilde{\mathcal{L}}_{\lambda}^{#1})}

\begin{document}
%
\title{$\ell_p$-Regularized Least Squares $(0<p<1)$
and\\[.5em] Critical Path}
%
%

\author{~\\[6em]
Masahiro~Yukawa,\\
Keio University, Dept.~Electronics and Electrical
        Engineering,\\
Hiyoshi 3-14-1, Kohoku-ku, Yokohama, Kanagawa, 223-8522 JAPAN\\
yukawa@elec.keio.ac.jp\\[5em]
        Shun-ichi~Amari,\\
RIKEN Brain Science Institute, the Laboratory for Mathematical
Neuroscience,\\
Hirosawa 2-1, Wako, Saitama, 351-0106 JAPAN\\
amari@brain.riken.jp\\[6em]
\thanks{This work was partially supported by a JSPS Grant-in-Aid
        (24760292). 
A preliminary version of this work was presented at the IEEE
International Symposium on Information Theory (ISIT), 2012 \cite{yukawa_isit12}.} 
}

\markboth{}%
{}
%



\maketitle







\begin{abstract}
The least squares problem is formulated in terms of
$\ell_p$ quasi-norm regularization ($0<p<1$).
Two formulations are considered:
(i) an $\ell_p$-constrained optimization and 
(ii) an $\ell_p$-penalized (unconstrained) optimization.
Due to the nonconvexity of the $\ell_p$ quasi-norm,
the solution paths of the regularized least squares problem
are not ensured to be continuous.
{\it A critical path}, which is a maximal continuous curve
consisting of critical points,
is therefore considered separately.
The critical paths are piecewise smooth, 
as can be seen from the viewpoint of the variational method,
and generally contain non-optimal points such as
saddle points and local maxima as well as global/local minima.
Along each critical path,
the correspondence between the regularization parameters
(which govern the 'strength' of regularization in the two formulations)
is non-monotonic and, more specifically, it has multiplicity.
Two paths of critical points connecting the origin and an ordinary least
squares (OLS) solution are highlighted.
One is a main path starting at an OLS solution, and the other is a greedy path
starting at the origin.
Part of the greedy path can be constructed with
a generalized Minkowskian gradient.
The breakpoints of the greedy path
coincide with the step-by-step solutions generated by
using orthogonal matching pursuit (OMP),
thereby establishing a direct link between OMP and
$\ell_p$-regularized least squares.
\end{abstract}


%

\clearpage

\section{Introduction}\label{sec:intro}

The present paper addresses the least squares problem 
by giving two different formulations for the $\ell_p$ quasi-norm ($0<p<1$)
regularization.
We will use a simple linear system model:
\begin{equation}
 \signal{y}:=[y_1, y_2, \cdots, y_d]^{\sf T}
= \signal{X}^{\sf T}\signal{\beta}_o + \signal{v}\in\real^d,
\end{equation}
where
$\signal{X}:=[\signal{x}_1 \ \signal{x}_2 \ \cdots
\signal{x}_n]^{\sf T}
\in\real^{n\times d}$ is a known matrix
with its columns being the design variables, 
$\signal{\beta}_o\in\real^{n}$ consists of the (unknown) explanatory
parameters, and $\signal{v}\in\real^d$ is the noise vector.
The first formulation under $\ell_p$-regularization for $p>0$
is as follows:\footnote{
The formulation \refeq{eq:c_min_problem} is essentially equivalent to
the following problem:
$\Rep{p}~ {\rm minimize}_{\signal{\beta}\in\real^n}
F_p(\signal{\beta})$
subject to
$\varphi(\signal{\beta})\leq \varepsilon$ for a given $\varepsilon\geq 0$.
Problem $\Rep{p}$, and thus problem $\Pc{p}$, for $0<p\leq 1$
is a relaxation of the sparse
optimization problem:
$\Rep{0}~{\rm minimize}_{\signal{\beta}\in\real^n}
\norm{\signal{\beta}}_0$
subject to
$\varphi(\signal{\beta})\leq \varepsilon$.
Here $\norm{\cdot}_0$ counts the number of nonzero entries of a vector.
}
\begin{align}
\hspace*{-1em}\Pc{p} &~~
{\rm minimize}_{\signal{\beta}\in\real^n}
~ \varphi(\signal{\beta}):=\frac{1}{2}
\norm{\signal{X}^{\sf T}\signal{\beta}-\signal{y}}_2^2
\nonumber\\
\hspace*{-1em}& \mbox{~ subject to }
F_p(\signal{\beta}) :=
\frac{1}{p} \norm{\signal{\beta}}_p^p
=  \sum_{i=1}^{n} \psi_p(\beta_i)\leq c,
\label{eq:c_min_problem}
\end{align}
where $c\geq 0$, 
$\norm{\cdot}_p$ denotes the $\ell_p$ (quasi-)norm for any $p>0$,
and $\psi_p(\beta):=\frac{1}{p}\abs{\beta}^p$, $\beta\in\real$.
Problem $\Pc{p}$ is referred to
as {\it the $\ell_p$-constrained least squares problem}.
The second formulation is as follows:
\begin{equation}
\Ql{p} ~~~
{\rm minimize}_{\signal{\beta}\in\real^n} ~~~
f_{\lambda}(\signal{\beta}):=\varphi(\signal{\beta})  +
\lambda F_p(\signal{\beta}),
\label{eq:lambda_min_problem}
\end{equation}
where $\lambda\geq 0$ is a Lagrange multiplier.
Problem $\Ql{p}$ is referred to as {\it the $\ell_p$-penalized least
squares problem}.

Both problems for $p\leq 1$
are related closely to 
sparse optimization problems encountered in various applications
and have therefore been studied extensively.
In the context of sparse signal recovery or compressed sensing
\cite{chen98,candes08,elad_book10},
underdetermined systems ($n\gg d$) are assumed, and the object is to
recover a sparse unknown vector from a small number of measurements.
In the context of model selection \cite{tibshirani96,efron04},
it is desired to select variables based on a sufficiently large number
(or sometimes a small number) of measurements.
In the case of $p=1$,
$F_1$ (i.e., the $\ell_1$ norm) is a convex function, and
it is widely known that $\Pc{1}$ and $\Ql{1}$ are equivalent in the sense
that the solutions of these problems coincide to each other
(and also that there is a continuous monotone correspondence between $c$ and $\lambda$).
In this case, $\Pc{1}$ is referred to as a {\it Lasso} \cite{tibshirani96}.
The least angle regression (LARS) algorithm
has been proposed \cite{efron04}
for constructing the solution path of $\Pc{1}$ with the value of $c$
sliding from zero to infinity.
Although LARS has been mainly studied in connection with overdetermined systems 
\cite{efron04},
it has also been applied to underdetermined systems (see \cite{donoho08}).

The $\ell_p$ norm becomes closer to the $\ell_0$ norm
as $p$ approaches zero, although $F_p$ is a nonconvex function
for $p<1$.
Considerable effort has therefore been devoted to the least squares
problem formulated in terms of $\ell_p$ norm regularization for $p<1$
\cite{gribonval03,elad07,chartrand08,davies09,foucart09,chen10,wang11,lai11,xu12}.
It has been shown experimentally that
the use of the $\ell_p$ norm yields
a sparser solution and a lower prediction error for model selection
compared with the $\ell_1$ norm  \cite{xu12}.
It has also been proven that fewer measurements as well as weaker
conditions are enough for sparse signal recovery
\cite{chartrand08,xu12,lai11}.
It is, therefore, important to see whether equivalence between $\Pc{p}$
and $\Ql{p}$ holds even for $p<1$ and, if not, how the equivalence is
modified.
As yet however, this fundamental question has not been
investigated.

In this paper, we shed light on this hitherto uninvestigated question
through an extension of LARS to the nonconvex case of $p<1$.
As expected, the case of $p<1$ is significantly different from the case
of $p=1$ due to the nonconvexity of $F_p$.
We prove that the solutions (i.e., the global minima) of $\Pc{p}$ and
$\Ql{p}$ are different for $p<1$.
However, there is a remarkable correspondence between the critical points
of $\Pc{p}$ and $\Ql{p}$.
The present paper studies the critical paths of the two problems and
elucidates their structures.
The main body of the paper consists of three parts.
In the first part, we study the solution paths
(the paths of global minima) of the problems $\Pc{p}$ and $\Ql{p}$
with the parameters $c$ and $\lambda$ sliding from zero to infinity
and show that the two paths are different from each other.
The solution of $\Pc{p}$ for $c=0$ is obviously the zero vector, and
as $c$ increases continuously, the solution moves away from
the origin continuously.
Indeed, the behavior of the $\Pc{p}$ solution path in the vicinity of
the origin is homotopically the same as that of the $\Pc{1}$ solution path.
On the other hand, the $\Ql{p}$ path is quite different.
The solution of $\Ql{p}$ for a sufficiently large $\lambda$ is
the zero vector and, as $\lambda$ decreases continuously, 
the solution jumps from the origin to a point on the $\Pc{p}$ path.
In short, the $\Ql{p}$ path is always {\it discontinuous}
at the origin,
whereas the $\Pc{p}$ path always leaves the origin continuously.
(Note, however, that the continuity of the whole $\Pc{p}$ path
is not necessarily guaranteed,
as will be seen in Example \ref{example:p07} of the appendix.)
In addition, the positive semi-definiteness of the Hessian matrix
of $f_{\lambda}$ is a necessary and sufficient condition for 
local minimality in $\Ql{p}$, but
it is only sufficient for local minimality in $\Pc{p}$.
As a result, the $\Pc{p}$ path contains the $\Ql{p}$ path
as its proper subset.

In the second part, we enlarge the problems to fill the gap by studying
the paths of {\it critical points} of
$\Pc{p}$ and $\Ql{p}$, which include local minima/maxima and saddle
points.
Strictly speaking, we address the following pair of problems:
\begin{align}
\Pct{p} & ~~~ \mbox{find critical points of }\Pc{p};\\
\Qlt{p} & ~~~ \mbox{find critical points of }\Ql{p}.
\end{align}
Critical points are defined by the first-order condition 
in their neighborhoods.
There are in general multiple critical points corresponding to
each value of $c$ or $\lambda$.
A critical point can therefore be regarded as a multiple-valued function
of $c$ (or $\lambda$).
We divide the set of all critical points into a smallest number of
subsets each of which forms a continuous curve in $\real^n$
that is a single-valued function of $c$ (or $\lambda$).
We call each of these curves
{\it a critical path of $\Pct{p}$ (or $\Qlt{p}$)}, or simply
{\it a $\Pct{p}$ path  (or an $\Qlt{p}$ path)} for short.
A remarkable difference from the case of $p=1$ is that
the correspondence between $c$ and $\lambda$ has multiplicity;
a single value of $\lambda$ corresponds to multiple values of $c$.
A critical path is a piecewise smooth curve and 
its smooth segments are characterized by a differential equation in $\real^n$.
The support of a critical point changes at each {\it breakpoint} at
which the direction of the curve changes discontinuously.
(A breakpoint is indeed a connection point of smooth curves in a critical path.)
At any breakpoint, (i) $\lambda=0$ and (ii)
the partial derivative of $\varphi$ with respect to every nonzero
component of $\signal{\beta}$ is zero.
We analyze the critical paths based on the variational method and
present {\it the connection theorem} that states that
two curves touch tangentially at the breakpoint connecting them.

In the third part, we study two paths of critical points
connecting the origin and an ordinary least squares (OLS) solution:
{\it a main path} and {\it a greedy path}.
A main path starts from an OLS solution and 
the active indices become inactive at breakpoints one by one.
A greedy path, on the other hand, 
starts from the origin and
indices become active at breakpoints one by one.
A simple modification can make the greedy path coincide
with the main path.
Part of the greedy path, 
on which the Hessian matrix is positive semidefinite,
can be constructed with a generalized Minkowskian gradient.
Both paths are composed of a union of critical paths, and hence
are piecewise smooth curves.
The breakpoints of the greedy path coincide exactly with the step-by-step
solutions generated by orthogonal matching pursuit (OMP) and thus,
bridge OMP\cite{tropp_omp07} and $\ell_p$-regularized least squares
problems.
This link is more direct than the one between OMP and the $\ell_1$
minimization established in \cite{donoho08}.

\section{Global Solution Paths}
\label{sec:analysis1}

In this section, we study the solution paths of $\Pc{p}$ and $\Ql{p}$
with $c$ and $\lambda$ in \refeq{eq:c_min_problem} and
\refeq{eq:lambda_min_problem}, respectively,
sliding from zero to infinity.
We refer to the paths simply as the $\Pc{p}$-path and $\Ql{p}$-path.
It is readily verified that
\begin{equation}
\varphi(\signal{\beta})=
\frac{1}{2}(\signal{\beta} - \signal{\beta}^*)^{\sf T}
\signal{G}(\signal{\beta} - \signal{\beta}^*)
+\gamma, ~~~\signal{\beta}\in\real^n,
\end{equation}
where $\signal{G}:=\signal{X}\signal{X}^{\sf T}$,
$\gamma:=\norm{\signal{y}}_2^2
- {\signal{\beta}^*}^{\sf T} \signal{G}\signal{\beta}^*$
is a constant in $\signal{\beta}$, and
\begin{equation}
 \signal{\beta}^*:=[\beta_1^*,\beta_2^*,\cdots,\beta_n^*]^{\sf T}
\in V^*:=
\argmin_{\signal{\beta}\in\real^n}\varphi(\signal{\beta}) 
\end{equation}
is an OLS solution.
In particular,
$\signal{\beta}^*:=(\signal{X}^{\sf T})^{\dagger}\signal{y}$
with the Moore-Penrose pseudo-inverse $(\signal{X}^{\sf T})^{\dagger}$
has the minimum norm among all the OLS solutions.

\subsection{Global Minimum}
\label{subsec:main_and_examples}

We denote by $\signal{\beta}_c^*$ and $\signal{\beta}_{\lambda}^*$
the global minima of $\Pc{p}$ and $\Ql{p}$
for given $c$ and $\lambda$, respectively.
In the case of $p\geq 1$, the following facts are well-known.
\clearpage
\begin{fact}[For $p\geq 1$]
\label{fact:l1}
~
\begin{description}
 \item[(a)] $\Pc{p}$ and $\Ql{p}$ are convex problems.
 \item[(b)] The $\Pc{p}$-path is unique.
 \item[(c)] The $\Ql{p}$-path is unique.
 \item[(d)] The correspondence between the solutions of $\Pc{p}$ and
       $\Ql{p}$ is one to one, and $\lambda$
is a continuous, monotone-decreasing, and single-valued function of $c\in(0,c^*)$,
where $c^*$ is the minimum value of $F_p$ among all the OLS solutions.
\end{description} 
\end{fact}

In the present case of $p< 1$, however, there are remarkable differences
between the two problems.
\begin{fact}[For $p< 1$]
\label{fact:lp}
 $\Pc{p}$ and $\Ql{p}$ are nonconvex problems and 
local minima exist in general.
\end{fact}
\begin{theorem}[Relation between $\Ql{p}$ and $\Pc{p}$ paths]
\label{theorem:proper_subset}
For $p< 1$,  the $\Ql{p}$-path is a proper subset of the $\Pc{p}$-path.
\end{theorem}

Theorem \ref{theorem:proper_subset} is 
the main result of this section, and it indicates an intrinsic difference between
$\Pc{p}$ and $\Ql{p}$.
Before proving it, we present a very simple example
to facilitate understanding of the theorem.

\begin{figure}[t]
\centering
\begin{psfrags}
\psfrag{c}[c][1.0]{$c$}
\psfrag{b}[c][1.0]{$\beta_{c}^*$}
\begin{center}
 \includegraphics[width=10cm]{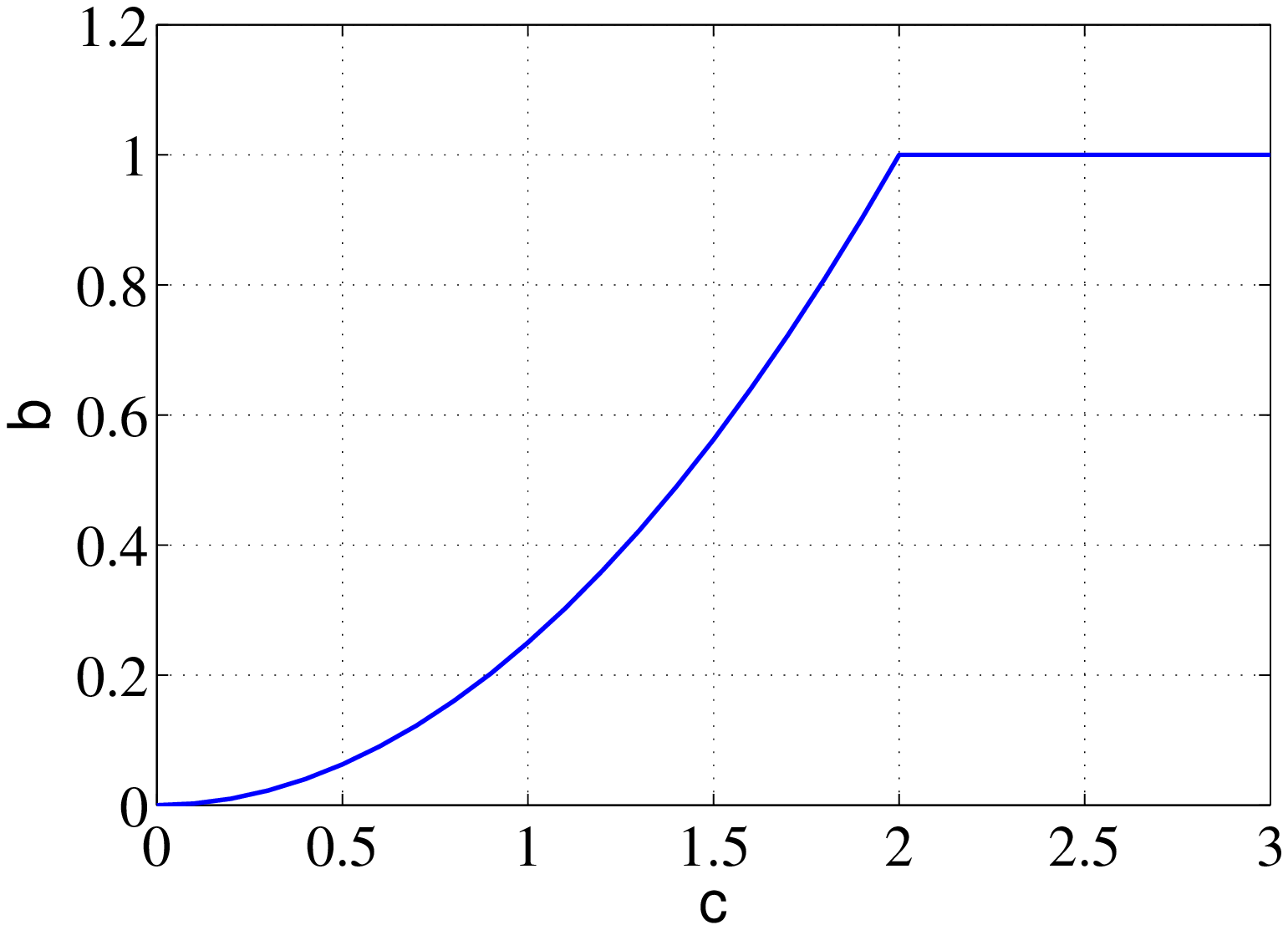} 
\end{center}
\end{psfrags}
\caption{Relation between $c$ and $\signal{\beta}_c^*$ 
 in Example \ref{example:1D}.}
\label{fig:c_beta_1D}
\end{figure}

\begin{figure}[t]
\centering
\begin{psfrags}
\psfrag{l}[c][1.0]{$\lambda$}
\psfrag{ltl}[c][1.0]{$\lambda_{\rm gl}$}
\psfrag{b}[c][1.0]{$\beta_{\lambda}^*$}
\psfrag{bgl}[c][1.0]{$\beta_{\rm gl}$}
\psfrag{bcr}[c][1.0]{$\beta_{\rm cr}$}
\begin{center}
 \includegraphics[width=10cm]{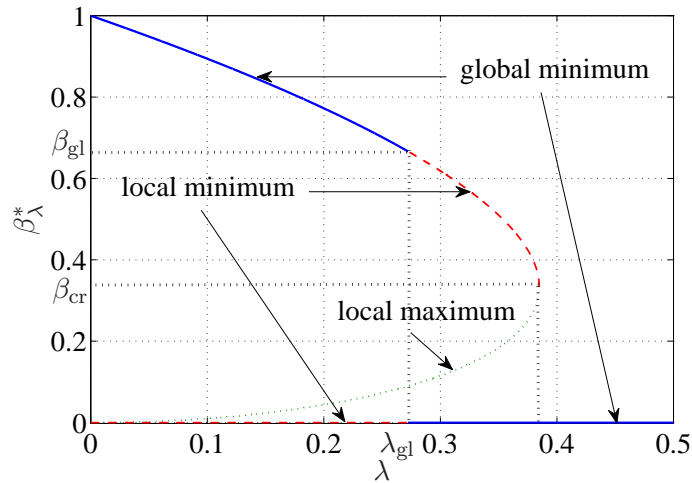} 
\end{center}
\end{psfrags}
\caption{Relation between $\lambda$ and $\signal{\beta}_{\lambda}^*$
 in Example \ref{example:1D}.}
\label{fig:lambda_beta_1D}
\end{figure}

\begin{example}[Global solution paths for 1D case]
\label{example:1D}
Consider the following one-dimensional problem ($p=0.5$):
$\varphi(\beta):=\frac{1}{2}(\beta-1)^2$  and
$F_{0.5}(\beta):=2|\beta|^{0.5}$.
It is clear that the solution $\beta_c^*$ of $\Pc{0.5}$
continuously changes from $\beta=0$ to the minimum $\beta^*=1$
of $\varphi(\beta)$ as $c$ increases from $c=0$ to
$c^*:=F_{0.5}(\beta^*)=2$,
and stays at $\beta=\beta^*$ as it increases beyond $c^*$.
The solution path of $\Pc{0.5}$ is, therefore,
the interval $[0,1]$ (see Fig.~\ref{fig:c_beta_1D}).
In contrast, the solution of $\Ql{0.5}$ changes {\it discontinuously}
at the origin, as will be shown below.

Figure \ref{fig:discontinuity} illustrates
the graphs of the cost function $f_{\lambda}(\signal{\beta})$ in
 \refeq{eq:lambda_min_problem} for different values of $\lambda$.
Looking at the red curve, which corresponds to $\lambda=0.3$, we can see that
there is a pair of local minima (the first one at the origin and the
second one between 0.5 and 1) and a single local maximum between 0 and 0.5.
As $\lambda$ decreases from $\lambda=0.3$ gradually to zero, 
the second local minimum approaches $\beta^*(=1)$
and the local maximum approaches the origin while the first local
minimum stays at the origin.
Increasing $\lambda$, on the other hand, 
the second local minimum and the local maximum approach each other
and merge into a single inflection point at $\lambda=0.385$
(the green curve).
As $\lambda$ increases beyond $0.385$,
$f_\lambda$ becomes a monotonically increasing function over
$[0,\infty)$
(the blue curve which corresponds to $\lambda=0.5$).
Therefore, there is a single local minimum at the origin, 
which is a sole critical point, and no local maximum
for $\lambda>0.385$.
Let us now consider how the solution (i.e., the global minimum) changes
depending on $\lambda$.
Starting from a large $\lambda$, we decrease it gradually.
The solution stays at the origin until $\lambda=0.385$.
For $\lambda$ slightly smaller than $0.385$,
the global minimum still stays at the origin,
since the value at the origin (the first local minimum) is smaller than the one 
at the second local minimum, as in the case of $\lambda=0.3$.
However, as $\lambda$ decreases further, 
$f_{\lambda}(\signal{\beta})$ at the second local minimum
decreases, while $f_{\lambda}(\signal{0})= 0.5$ for any $\lambda\geq 0$.
The second local minimum eventually becomes a global minimum at some
value, say $\lambda_{\rm gl}$\hspace*{.1em}, between 0.2 and 0.3.
This implies that the solution of $\Ql{0.5}$ jumps from $\beta=0$ to 
$\beta_{\rm gl}\in\real$\hspace*{.1em}, which is a global
minimum of $f_{\lambda_{\rm gl}}$ satisfying
$f_{\lambda_{\rm gl}}(0)=f_{\lambda_{\rm gl}}(\beta_{\rm gl})$.
As $\lambda$ decreases from $\lambda_{\rm gl}$ to zero,
the solution changes from $\beta_{\rm gl}$ to $\beta^*(=1)$.
The solution path of $\Ql{0.5}$ thus consists of disjoint sets
$\{0\}\cup [\beta_{\rm gl},\beta^*]$
(see Fig.~\ref{fig:lambda_beta_1D}).
Figure \ref{fig:lambda_beta_1D} will be discussed
later in Example \ref{example:1D_critical_points}.
\end{example}

\subsection{Local Optimality in $\Pc{p}$ and $\Ql{p}$}
\label{subsec:local_opt}

Apart from the global minimum, let us examine the conditions for local minimality
in $\Pc{p}$ and $\Ql{p}$.
Lemma \ref{lemma:ns} below shows that 
$\Pc{p}$ and $\Ql{p}$ have different local-minimality characteristics.
In $\Pc{p}$, a point $\signal{\beta}$ is a local minimum
when the function $\varphi$ is locally minimal over
the (nonconvex) constraint set 
$\mathcal{B}_c:=\{\signal{\beta}\in\real^n :F_p(\signal{\beta}) \leq c \}$.
In $\Ql{p}$, on the other hand, a point $\signal{\beta}$ is a local minimum
when the function $\varphi+\lambda F_p$ is locally minimal
over the whole Euclidean space $\real^n$.
In short, local minimality in $\Pc{p}$ is defined as
that of the convex function over the nonconvex constraint set
$\mathcal{B}_c$, 
whereas local minimality in $\Ql{p}$ is defined as that of the
nonconvex function without any constraint.
This makes an essential difference between the local minimality conditions
for $\Pc{p}$ and $\Ql{p}$.

\begin{figure}[t]
\centering
\psfrag{Bc}[c][1.0]{$\mathcal{B}_c$}
\psfrag{R}[c][1.0]{$\mathcal{R}$}
\psfrag{b}[c][1.0]{$\hat{\signal{\beta}}$}
 \includegraphics[width=8cm]{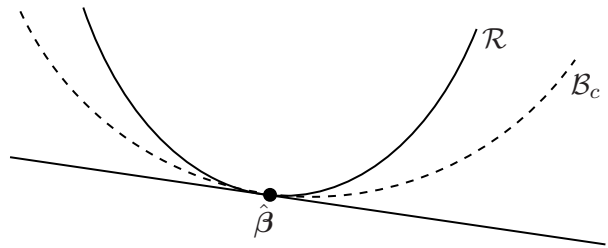}
\caption{Case that $\hat{\signal{\beta}}$ is a local minimum of ($\mathcal{P}_c^p$).}
\label{fig:local_min}
\end{figure}

We can geometrically describe local minimality of a point
$\hat{\signal{\beta}}$ in $\Pc{p}$ as follows.
Let $\mathcal{R}$ denote the contour of the function $\varphi$ passing 
through the point $\hat{\signal{\beta}}$.
Also, let $\partial {\mathcal B}_c$ denote the boundary of ${\mathcal B}_c$ for
$c:=F_p(\hat{\signal{\beta}})$.
Suppose for simplicity that there exists a unique OLS solution
$\signal{\beta}^*:=(\signal{X}^{\sf T})^{\dagger}\signal{y}$;
i.e., $\varphi$ is strictly convex and the problem is overdetermined.
To exclude trivial cases,
we will assume that  $\signal{\beta}^*$
(the center of $\mathcal{R}$) is located outside the constraint
set ${\mathcal B}_c$. 
Suppose that $\hat{\signal{\beta}}$ has no zero components.
In this case, $\hat{\signal{\beta}}$ is a local minimum
if (i) the two surfaces $\mathcal{R}$ and $\partial {\mathcal B}_c$ 
touch each other (i.e., share the same tangent plane)
at $\hat{\signal{\beta}}$, and 
(ii) $\partial {\mathcal B}_c$ is closer to the tangent
plane than $\mathcal{R}$ in the vicinity of $\hat{\signal{\beta}}$
(see Fig.~\ref{fig:local_min}).
In the case that $\hat{\signal{\beta}}$ has some zero components,
the above geometric properties hold in the subspace where
zero-components of $\hat{\signal{\beta}}$ are fixed to zero.

Given any vector
$\signal{\beta}:=[\beta_1,\beta_2,\cdots,\beta_n]^{\sf T}\in\real^n$,
we define the set of its active indices as
${\rm supp}(\signal{\beta}):=\{i\in\{1,2,\cdots,n\}:
\beta_i\neq 0\}$.
Let $\mathcal{I}:=\{i_1,i_2,\cdots,i_s\}:={\rm supp}(\signal{\beta})$,
where $s:=\abs{{\rm supp}(\signal{\beta})}$ means
the cardinality of 
${\rm supp}(\signal{\beta})$; i.e., 
$\signal{\beta}$ is supposed to have $s$ nonzero entries
$\beta_{i_1}$, $\beta_{i_2}$, $\cdots$, $\beta_{i_s}\neq 0$.
Define a sub-vector 
$\signal{\beta}_{\mathcal{I}}:=[\beta_{i_1},\beta_{i_2},\cdots,
\beta_{i_s}]^{\sf T}\in\real^s$ of
$\signal{\beta}$ consisting of its nonzero components.
We denote by $\signal{\nabla}_\mathcal{I}$ the gradient in terms of
the nonzero components; e.g.,
\[
 \signal{\nabla}_\mathcal{I}\varphi(\signal{\beta}):= 
\left[
\partial_{i_1}\varphi(\signal{\beta}),
\partial_{i_2} \varphi(\signal{\beta}),
\cdots,
\partial_{i_s} \varphi(\signal{\beta})
\right]^{\sf T},
\]
where the simplified notation $\partial_{i}$
is used
rather than $\partial/\partial \beta_i$,
to denote the partial derivative with respect to $\beta_i$.
The first and second derivatives of
$\psi_p(\beta)(:= \frac{1}{p}\abs{\beta}^p)$ at a point $\beta\neq 0$
are, respectively, given by
\begin{align}
\psi_p'(\beta) = & \
\mbox{sgn}(\beta)\abs{\beta}^{-(1-p)},
\label{eq:nabla_F1half}\\
\psi_p''(\beta) =  & \
-(1-p)\abs{\beta}^{-(2-p)},
\label{eq:nabla2_F1half}
\end{align}
where $\mbox{sgn}(\cdot)$ is the signum function.
The following lemma presents necessary and sufficient conditions 
for local minimality in $\Pc{p}$ and $\Ql{p}$.

\begin{lemma}
[Necessary and sufficient conditions for local minimality in $\Pc{p}$ and $\Ql{p}$]
\label{lemma:ns}
~\\[-1.5em]
\begin{enumerate}
 \item A vector $\hat{\signal{\beta}}$ is a local minimum
of $\Ql{p}$ if, and only if,
(i) it satisfies the first-order condition,
\begin{equation}
\signal{\nabla}_\mathcal{I}\varphi(\hat{\signal{\beta}})
= - \lambda \signal{\nabla}_\mathcal{I} F_p(\hat{\signal{\beta}}),
\label{eq:necessary_sufficient_condition_Rl}
\end{equation}
where $\mathcal{I}:={\rm supp}(\hat{\signal{\beta}})$,
and (ii) the Hessian matrix,
\begin{equation}
 \signal{K}(\hat{\signal{\beta}}):=
\signal{\nabla}_\mathcal{I} \signal{\nabla}_\mathcal{I} 
\left(
\varphi + \lambda F_p
\right)(\hat{\signal{\beta}})
\end{equation}
is positive semidefinite.

\item A vector $\hat{\signal{\beta}}$ is a local minimum
of $\Pc{p}$ if, and only if,
(i) it satisfies the first-order condition,
\begin{equation}
\signal{\nabla}_\mathcal{I}\varphi(\hat{\signal{\beta}})
= - \lambda_c \signal{\nabla}_\mathcal{I} F_p(\hat{\signal{\beta}})
\label{eq:first_order_condition}
\end{equation}
for some $\lambda_c \geq 0$, where $\mathcal{I}:={\rm supp}(\hat{\signal{\beta}})$,
and (ii) the Hessian matrix
$ \signal{K}(\hat{\signal{\beta}})$ with $\lambda:=\lambda_c$
is either positive semidefinite (for all vectors) or
positive definite
for any tangent vector $\signal{\epsilon}$ of the contour of
      $F_p$ passing through $\hat{\signal{\beta}}$;
i.e., 
$\signal{x}^{\sf T} \signal{K}(\hat{\signal{\beta}})\signal{x}\geq 0$
for all $\signal{x}\in\real^{\abs{\mathcal{I}}}$, or
 $\signal{\epsilon}^{\sf T}
\signal{K}(\hat{\signal{\beta}})
\signal{\epsilon} > 0$
for all $\signal{\epsilon}\neq \signal{0}$ satisfying
$\signal{\nabla}_\mathcal{I}F_p(\hat{\signal{\beta}})^{\sf T}
\signal{\epsilon}=0$.
\end{enumerate}
\end{lemma}
\noindent Proof:
Lemma \ref{lemma:ns}.1 is clear.
We prove
Lemma \ref{lemma:ns}.2 as follows.
Although the statement is true for an arbitrary $\mathcal{I}$,
we only provide a proof for the case that $\mathcal{I}=\{1,2,\cdots,n\}$.
We drop the index $\mathcal{I}$ for simplicity.
The first part is a condition for $\hat{\signal{\beta}}$
to be a critical point.
Noting that every local minimum, say $\tilde{\signal{\beta}}$, satisfies
$F_p(\tilde{\signal{\beta}})\leq F_p(\signal{\beta}^*)$,
$\hat{\signal{\beta}}$ is a local minimum
if, and only if, there exists a $\delta>0$ such that
\begin{equation}
 \varphi(\hat{\signal{\beta}} + \Delta \signal{\beta}) \geq
 \varphi(\hat{\signal{\beta}})
\label{eq:original_sufficiency}
\end{equation}
for any $\Delta \signal{\beta}\in\real^n$ satisfying
\begin{align}
F_p(\hat{\signal{\beta}} + \Delta \signal{\beta}) = & \
F_p(\hat{\signal{\beta}}), \label{eq:epsilon_cond1}\\
 \norm{\Delta \signal{\beta}}_2 \leq  & \ \delta.
\label{eq:epsilon_cond2}
\end{align}
For a sufficiently small $\delta>0$,
Taylor expansions of $\varphi$ and $F_p$ are, respectively, given by
\begin{eqnarray}
\hspace*{-3em} 
& \varphi(\hat{\signal{\beta}} + \Delta \signal{\beta}) - 
 \varphi(\hat{\signal{\beta}}) =
\signal{\nabla} \varphi(\hat{\signal{\beta}})^{\sf T} \Delta \signal{\beta}
+ \dfrac{1}{2}\Delta \signal{\beta}^{\sf T} 
\signal{\nabla} \signal{\nabla} \varphi(\hat{\signal{\beta}})
\Delta \signal{\beta},
\label{eq:taylor_phi}\\
\hspace*{-2em}
& F_p(\hat{\signal{\beta}} + \Delta \signal{\beta}) - 
 F_p(\hat{\signal{\beta}}) = 
 \signal{\nabla}F_p(\hat{\signal{\beta}})^{\sf T} \Delta \signal{\beta}
+ \dfrac{1}{2}\Delta \signal{\beta}^{\sf T} 
\signal{\nabla} \signal{\nabla}F_p (\hat{\signal{\beta}})
\Delta \signal{\beta},
\label{eq:taylor_Fp}
\end{eqnarray}
where higher order terms have been neglected, and
$\Delta \signal{\beta}$ can be decomposed from \refeq{eq:epsilon_cond1} as
\begin{equation}
 \Delta\signal{\beta}=\nu\signal{\epsilon} + \alpha\signal{n},
~~~\nu>0, ~~~(0\leq)\alpha=o(\nu),
\label{eq:Delta_beta}
\end{equation}
where 
$\signal{\epsilon}$ and $\signal{n}$
denote a tangent vector and a normal vector
of the contour of $F_p$ passing through $\hat{\signal{\beta}}$,
respectively.
From \refeq{eq:first_order_condition},
\refeq{eq:epsilon_cond1}, and 
\refeq{eq:taylor_Fp}, we obtain
\begin{equation}
 \signal{\nabla} \varphi(\hat{\signal{\beta}})^{\sf T} \Delta\signal{\beta}
= - \lambda_c \signal{\nabla} F_p(\hat{\signal{\beta}})^{\sf T}
\Delta\signal{\beta}
= \frac{\lambda_c}{2}
\Delta\signal{\beta}^{\sf T} 
\signal{\nabla} \signal{\nabla}F_p (\hat{\signal{\beta}})
\Delta\signal{\beta},
\label{eq:gradphi_epsilon}
\end{equation}
which yields, together with \refeq{eq:taylor_phi} and
\refeq{eq:Delta_beta},
\begin{equation}
 \varphi(\hat{\signal{\beta}} + \Delta \signal{\beta}) - 
 \varphi(\hat{\signal{\beta}}) = 
\frac{1}{2}
\Delta\signal{\beta}^{\sf T} 
\signal{K}(\hat{\signal{\beta}})
\Delta\signal{\beta}
=\frac{\nu^2}{2}\signal{\epsilon}^{\sf T}\signal{K}(\hat{\signal{\beta}})\signal{\epsilon}
+ \underbrace{\nu\alpha
\signal{\epsilon}^{\sf T}\signal{K}(\hat{\signal{\beta}})\signal{n}
+ \frac{\alpha^2}{2}\signal{n}^{\sf T}
\signal{K}(\hat{\signal{\beta}})\signal{n}}_{\mbox{$=o(\nu^2)$}}.
\label{eq:diff_phi}
\end{equation}
This proves the second part.
A proof for an arbitrary $\mathcal{I}$ can be obtained by
noting that,
 due to \refeq{eq:epsilon_cond1}, 
the norm of
 $\Delta \signal{\beta}_{\bar{\mathcal{I}}}$, where 
$\bar{\mathcal{I}}:=\{1,2,\cdots,n\}\setminus \mathcal{I}$, 
diminishes quickly as $\delta$ approaches zero.
\migip

\begin{remark}
[Difference between $\Pc{p}$ and $\Ql{p}$ in terms of local minimality
 in Lemma \ref{lemma:ns}]
\label{remark:diff_Pc_Pl}
The positive semidefiniteness of the Hessian matrix
 $\signal{K}(\hat{\signal{\beta}})$ 
is a necessary and sufficient condition for $\Ql{p}$,
whereas it is only sufficient for $\Pc{p}$.
It is, therefore, possible that a vector $\hat{\signal{\beta}}$ is 
a local minimum of $\Pc{p}$, but a saddle of $\Ql{p}$,
as will be shown in Example \ref{example:2D} in Section  \ref{subsec:critical_path}.
Indeed, the RHS of {\rm \refeq{eq:diff_phi}} is positive
for a sufficiently small $\nu>0$ if
$\signal{\epsilon}^{\sf T}\signal{K}(\hat{\signal{\beta}})\signal{\epsilon}>0$
 (even if
$\signal{\epsilon}^{\sf T}\signal{K}(\hat{\signal{\beta}})\signal{n}$ and 
$\signal{n}^{\sf T}\signal{K}(\hat{\signal{\beta}})\signal{n}$ are
negative); i.e.,
$\signal{K}(\hat{\signal{\beta}})$ is allowed to be not positive
 semidefinite for a normal vector $\signal{n}$.
\end{remark}

\subsection{Proof of Theorem \ref{theorem:proper_subset}}

It is not difficult to see that the $\Ql{p}$-path is a subset of the
$\Pc{p}$-path.
The properness is verified by the following lemma
derived from Lemma \ref{lemma:ns}.

\begin{lemma}
\label{lemma:continuity}
For any $p\in(0,1)$,
\begin{enumerate}
 \item[(a)] $\Pc{p}$-path is continuous
at $\signal{\beta}=\signal{0}$;
 \item[(b)] $\Ql{p}$-path is discontinuous
at $\signal{\beta}=\signal{0}$.
\end{enumerate}
\migip
\end{lemma}

\noindent{\it Proof of Lemma \ref{lemma:continuity}:} \\
Proof of (a): 
It is clear that $\signal{\beta}(0)=\signal{0}$ 
since $\{\signal{\beta}:F_p(\signal{\beta})\leq 0\}=\{\signal{0}\}$
and that
$\norm{\signal{\beta}(c)-\signal{\beta}(0)}_2 = \norm{\signal{\beta}(c)}_2
\rightarrow 0$ as $c\rightarrow 0$, implying the continuity of
the $\Pc{p}$-path at the origin.

\noindent Proof of (b): 
Notice that $\varphi$ is differentiable over $\real^n$.
The function $F_p(\signal{\beta})$ can be expressed as
$F_p(\signal{\beta})=\sum_{i=1}^{n} \psi(\beta_i)$, where
$\psi(\beta):= \frac{1}{p}\abs{\beta}^p$ for $\beta\in\real$.
It can be verified that 
$\lim_{\beta\uparrow 0}\frac{d}{d\beta}\psi(\beta)= \infty$
and 
$\lim_{\beta\downarrow 0}\frac{d}{d\beta}\psi(\beta)= - \infty$.
This implies that $\signal{\beta}=\signal{0}$  is a local minimum 
of the function $\varphi(\signal{\beta})+ \lambda F_p(\signal{\beta})$
for any $\lambda>0$ and 
no local minima exist in a neighborhood of $\signal{\beta}=\signal{0}$.
Thus, the $\Ql{p}$-path is discontinuous
at $\signal{\beta}=\signal{0}$.
\migip

\section{Paths of Critical Point}
\label{sec:analysis2}

Section \ref{sec:analysis1}  showed that the $\Ql{p}$-path is always
discontinuous and is different from the $\Pc{p}$-path, which is
continuous at $\signal{\beta}=\signal{0}$.
It is beneficial to extend LARS to the nonconvex case of $p<1$
in such a way that the path is continuous.
Here, we extend the criterion from one of minimality to one of criticality
for the two problems, and consider continuous paths of critical points.
Although we denoted the dependency of $\lambda$ on $c$ by $\lambda_c$
in \refeq{eq:first_order_condition}, we will denote it by $\lambda(c)$
when viewing $\lambda$ as a function of $c$.
Similarly, we use the notation $c(\lambda)$.

\subsection{Critical point}
\label{subsec:critical_point}

The definition of critical points is as follows.
\begin{definition}
[Critical point]
\label{def:critical_point}
When $\widetilde{\signal{\beta}}\in\real^n$ satisfies
the first-order condition
\begin{equation}
\signal{\nabla}_\mathcal{I}\varphi(\widetilde{\signal{\beta}})
= -\widetilde{\lambda} \signal{\nabla}_\mathcal{I} F_p(\widetilde{\signal{\beta}})
\label{eq:necessary_condition}
\end{equation}
for some $\widetilde{\lambda}\geq 0$, 
where $\mathcal{I}:={\rm supp}(\widetilde{\signal{\beta}})$,
it is called a critical point of $\Pc{p}$ for $c:=F_p(\widetilde{\signal{\beta}})$,
or a critical point of $\Ql{p}$ for $\lambda:=\widetilde{\lambda}$.
\end{definition}
Note that condition \refeq{eq:necessary_condition} can be expressed
as follows:
\begin{equation}
 \frac{\partial_i
  \varphi(\widetilde{\signal{\beta}})}
{\partial_i F_p(\widetilde{\signal{\beta}})}
= \frac{\partial_i \varphi(\widetilde{\signal{\beta}})}
{ \psi'_p(\widetilde{\beta}_i)}
= - \widetilde{\lambda}, ~~~ \forall i\in\mathcal{I}, ~~~\exists
\widetilde{\lambda} \geq 0.
\end{equation}
Geometrically speaking,
$\widetilde{\signal{\beta}}$ is a critical point
when the two surfaces $\mathcal{R}$ and $\partial {\mathcal B}_c$
(see Section \ref{subsec:local_opt})
share the same tangent plane at $\widetilde{\signal{\beta}}$.
At a critical point $\widetilde{\signal{\beta}}$, the function $\varphi$ takes
a critical value over $\mathcal{B}_c$ for $c:=F_p(\widetilde{\signal{\beta}})$,
and, at the same time, the function $\varphi+\widetilde{\lambda}F_p$ takes a
critical value over $\real^n$.

\begin{proposition}
The following statements hold.
\begin{enumerate}
 \item A critical point of $\Pc{p}$ for any $c\geq 0$ is a critical
       point of $\Ql{p}$ for some $\lambda\geq 0$.

 \item A critical point of $\Ql{p}$ for any $\lambda\geq 0$ is a
       critical  point of $\Pc{p}$ for some
       $c\geq 0$.
\end{enumerate}
\end{proposition}

In the rest of this section, we consider problems
$\Pct{p}$ and $\Qlt{p}$ rather than $\Pc{p}$ and $\Ql{p}$.

\subsection{Critical path} \label{subsec:critical_path}

{\it The set of critical points} for $\Pct{p}$,
which is the same as that for $\Qlt{p}$, is given as
\begin{equation}
 \mathcal{C}:=\left\{
\widetilde{\signal{\beta}}\in\real^n : \mbox{there exists }\widetilde{\lambda}\geq 0
\mbox{ s.t.~\refeq{eq:necessary_condition} holds} 
\right\}.
\end{equation}
Some important observations are listed below.
\begin{enumerate}
 \item A local minimum of $\Ql{p}$ is a local minimum of $\Pc{p}$, but the
       converse is not true.
 \item The correspondence between $c$ and $\lambda(c)$ has multiplicity.
 \item The paths of the global minima of $\Pc{p}$ and $\Ql{p}$ are both subsets of
$\mathcal{C}$.
 \item The path of the global minima of $\Ql{p}$ is always discontinuous.
 \item The path of the global minima of $\Pc{p}$ is possibly discontinuous
(see Example \ref{example:p07} in the appendix).
\end{enumerate}

Each critical point $\widetilde{\signal{\beta}}$
is associated with a certain value of
$c~(=F_p(\widetilde{\signal{\beta}}))$, and in general,
there are multiple critical points that are
associated with a single value of $c$.
It is clear that
the origin is a unique critical point associated with
$c=0$.\footnote{In this case, the set of active indices
$\mathcal{I}$ is an empty set, and hence, the condition is automatically satisfied.}
As $c$ increases from zero, the multiple critical points
associated with each value of $c$ draw multiple curves in $\real^n$.
We call each such curve {\it a critical path of $\Pct{p}$}, which is
defined formally below.
Intuitively, a critical path is a maximal continuous curve that is a
single-valued function of $c$ (or $\lambda$).

\begin{definition}
[Critical path]
\label{def:critical_path}
~
\begin{enumerate}
 \item A subset $\tilde{\mathcal{C}}\subset\mathcal{C}$ is called
a critical path of $\Pct{p}$, or  a $\Pct{p}$ path for short,
if
(i) the mapping $T:  \tilde{\mathcal{C}} \rightarrow S\subset[0,\infty), \
 \widetilde{\signal{\beta}} \mapsto c=F_p(\widetilde{\signal{\beta}} )$ has
a one-to-one continuous inverse mapping $T^{-1}$, and
(ii) none of the proper supersets of $\tilde{\mathcal{C}}$ satisfies
       condition (i). 
 \item A subset $\tilde{\mathcal{C}}\subset\mathcal{C}$ is called
a critical path of $\Qlt{p}$, or a $\Qlt{p}$ path for short,
if
(i) the mapping $T: \tilde{\mathcal{C}}
 \rightarrow S\subset[0,\infty), \
\signal{0}\mapsto 0, \ 
 \widetilde{\signal{\beta}}(\neq \signal{0}) \mapsto \lambda=- 
\partial_i \varphi(\widetilde{\signal{\beta}})/
\partial_i F_p(\widetilde{\signal{\beta}})$,
       $i\in{\rm supp}(\widetilde{\signal{\beta}})$, has
a one-to-one continuous inverse mapping $T^{-1}$, and
(ii) none of the proper supersets of $\tilde{\mathcal{C}}$
satisfies condition (i).
\end{enumerate}
\end{definition}

Typical examples of critical paths are given below
to give the reader an intuitive understanding before the general analysis
of critical paths.

\begin{example}[Critical paths for 1D case]
 \label{example:1D_critical_points}
Consider the critical paths for the functions considered in Example
 \ref{example:1D}.
The function $f_{\lambda}$ has possibly three
 critical points: $\beta=0$ for any $\lambda> 0$ and
points $\widetilde{\beta}_{\lambda}$ included in the set
\begin{equation}
 R_{\lambda}(\beta^*)=
\left\{\beta>0:
f_{\lambda}'(\beta)= \beta-\beta^* +\frac{\lambda}{\sqrt{\beta}} = 0\right\}
\label{eq:critical_condition_1D}
\end{equation}
when $R_{\lambda}(\beta^*)\neq \emptyset$.
It can be verified that $\abs{R_{\lambda}(\beta^*)}=2$
for $\lambda<2\left(\beta^*/3\right)^{1.5}$ (see
 Fig.~\ref{fig:lambda_beta_1D}).
When three critical points exist for a $\lambda$, one is $\beta=0$.
The larger element of $R_{\lambda}(\beta^*)$ and $\beta=0$ are the local/global minima 
of $f_{\lambda}$; the other one is the local maximum,
as illustrated in Fig.~\ref{fig:lambda_beta_1D}.
While the $\Ql{0.5}$ global path
is the disjoint set $\{0\}\cup [\beta_{\rm gl},\beta^*]$
for a $\beta_{\rm gl}$,
the $\Qlt{0.5}$ critical paths are
two intervals: $[0,\beta_{\rm cr}]$
and $[\beta_{\rm cr},\beta^*]$ for a $\beta_{\rm cr}$.
In contrast, the $\Pct{0.5}$ critical path coincides with
the $\Pc{0.5}$ global path $[0,1]$ in this case,
although this is not always true.
\end{example}

\begin{figure}[t]
\centering
\psfrag{b}[c][1.0]{$\widetilde{\beta}_{\lambda}$}
\psfrag{l}[c][1.0]{$\lambda$}
 \includegraphics[width=8cm]{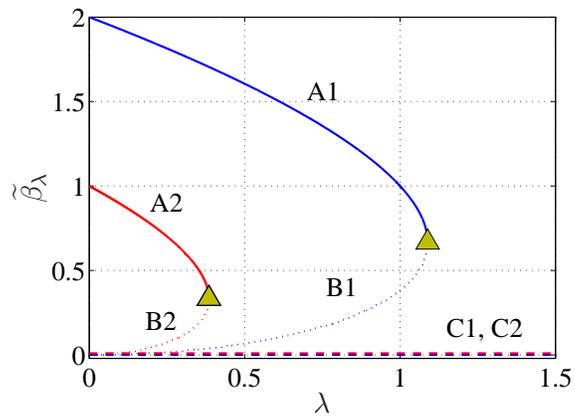}
\caption{Critical points of $f_{\lambda,1}(\beta_1)$ and $f_{\lambda,2}(\beta_2)$ in
 Example \ref{example:2D}.}
\label{fig:lambda_beta}
\end{figure}

\begin{figure}[t]
\centering
\begin{psfrags}
\subfigure[]{
\psfrag{b1}[c][1.0]{$\beta_1$}
\psfrag{b2}[c][1.0]{$\beta_2$}
 \includegraphics[width=8cm]{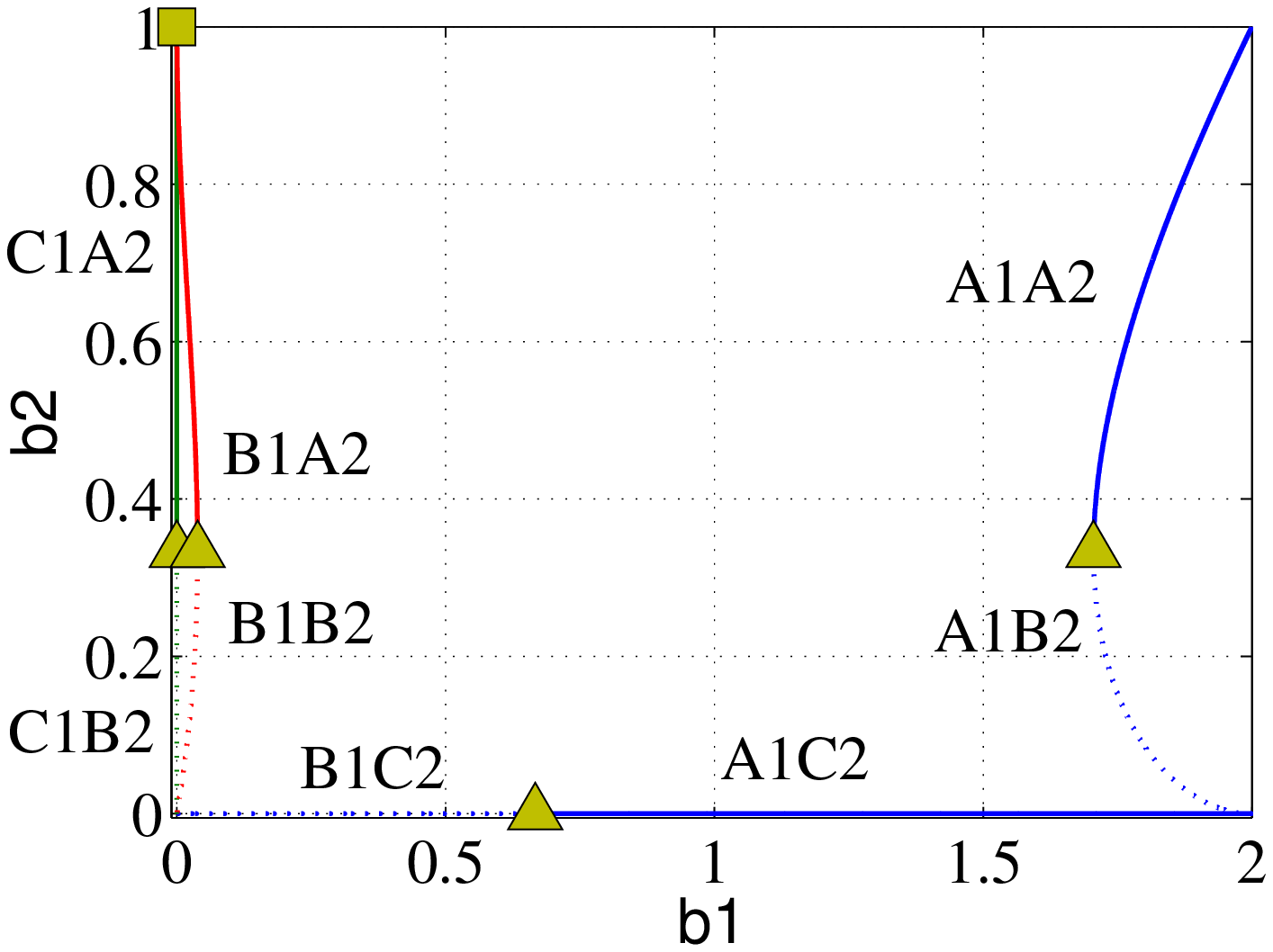}
}
\subfigure[]{
\begin{psfrags}
\psfrag{b}[c][1.0]{$\beta$}
\psfrag{c}[c][1.0]{$c$}
\psfrag{beta1path1}[c][1.0]{$\widetilde{\beta}_{c,1}$ (Path 1)}
\psfrag{beta2path1}[c][1.0]{$\widetilde{\beta}_{c,2}$ (Path 1)}
\psfrag{beta1path2}[c][1.0]{\hspace*{1em}$\widetilde{\beta}_{c,1}$ (Path 2)}
\psfrag{beta2path2}[c][1.0]{\hspace*{1em}$\widetilde{\beta}_{c,2}$ (Path 2)}
\psfrag{beta1path3}[c][1.0]{\hspace*{1em}$\widetilde{\beta}_{c,1}$ (Path 3)}
\psfrag{beta2path3}[c][1.0]{\hspace*{1em}$\widetilde{\beta}_{c,2}$ (Path 3)}
 \includegraphics[width=9cm]{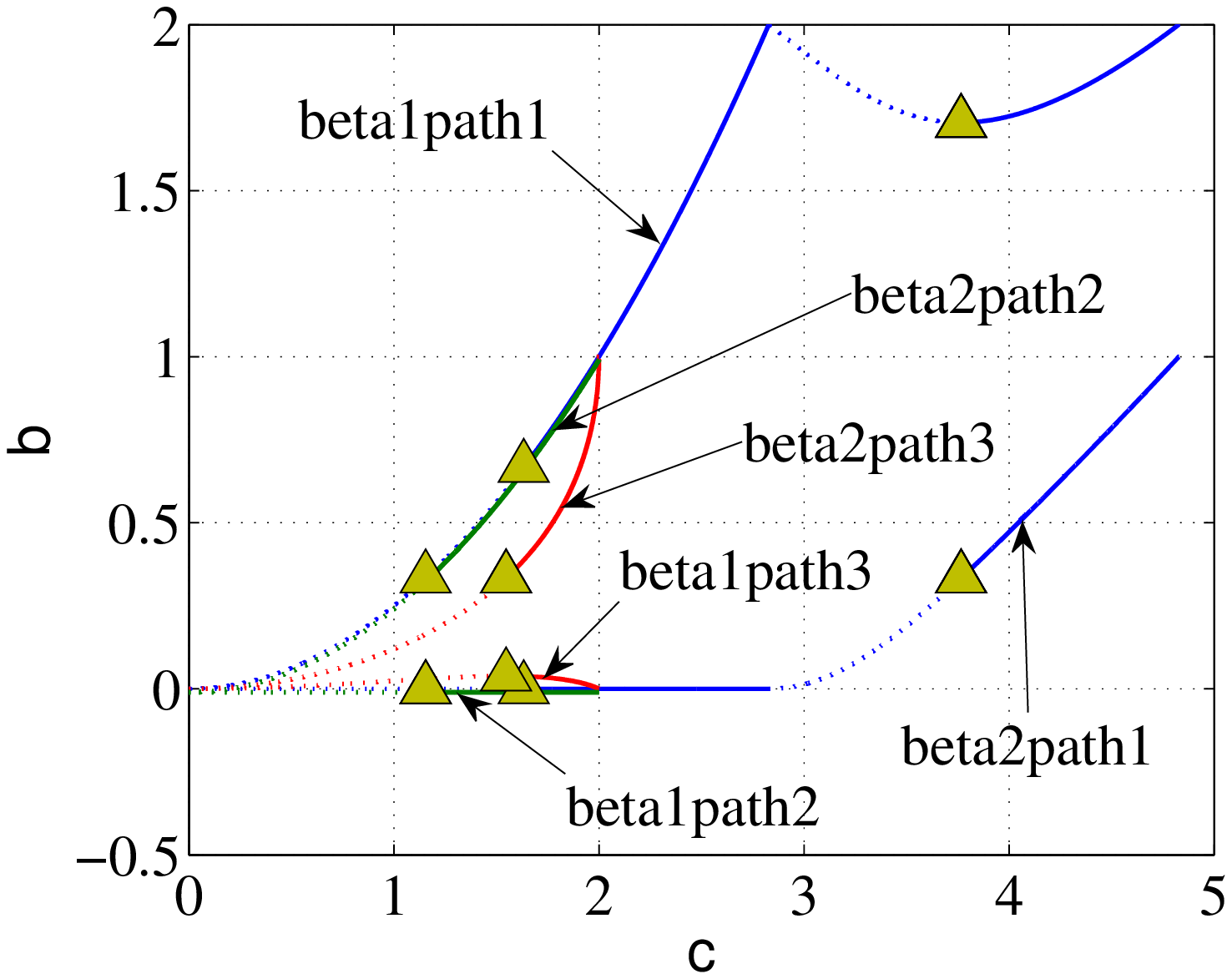}
\end{psfrags}
}
\subfigure[]{
\psfrag{c}[c][1.0]{$c$}
\psfrag{l}[c][1.0]{$\lambda(c)$}
 \includegraphics[width=8cm]{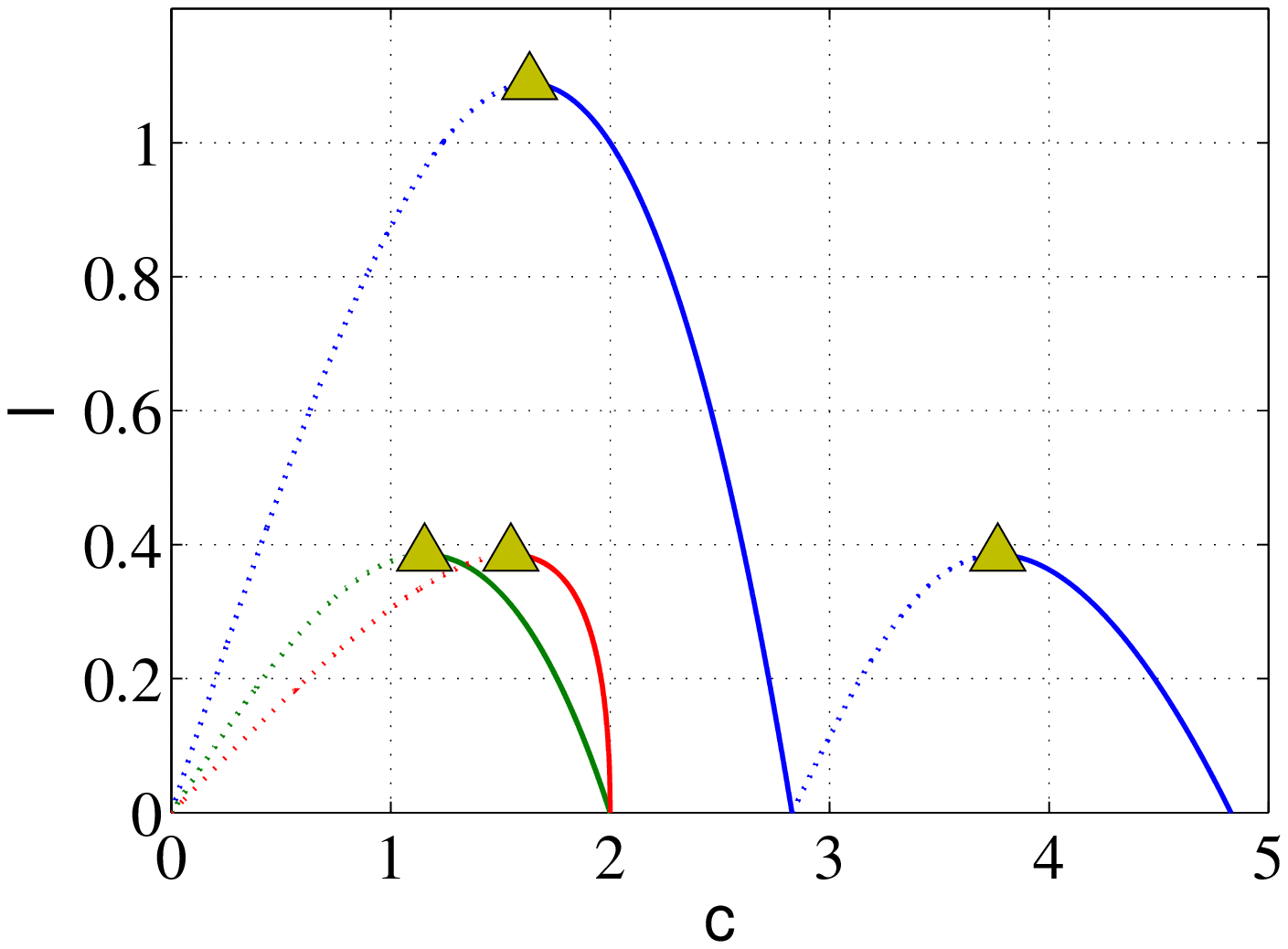}
}
\end{psfrags}
\caption{(a) Critical paths of $\Pct{0.5}$
with its correspondence to $\Qlt{0.5}$,
(b) critical point
$\widetilde{\signal{\beta}}_c=[\widetilde{\beta}_{c,1},
\widetilde{\beta}_{c,2}]^{\sf T}$
as a function of $c$,
and (c) $c$ - $\lambda(c)$ correspondence
in  Example \ref{example:2D}.}
\label{fig:c_trajectory}
\end{figure}

\begin{example}[2D orthogonal case]
\label{example:2D}
Consider the following two-dimensional case 
under the orthogonality condition
$\signal{G}:=\signal{X}\signal{X}^{\sf T}=\signal{I}$:
 $\varphi(\signal{\beta}):=\frac{1}{2}
\norm{\signal{\beta} - \signal{\beta}^*}_2^2$ with
$\signal{\beta}^*:=[2,1]^{\sf T}$
and
$F_{0.5}(\signal{\beta}):=2(|\beta_1|^{0.5} + |\beta_2|^{0.5} )$.
In this special case, $f_{\lambda}(\signal{\beta})$ can be decomposed as
\begin{equation}
f_{\lambda}(\signal{\beta})=
f_{\lambda,1}(\beta_1) + f_{\lambda,2}(\beta_2), 
\label{eq:f_lambda_separate}
\end{equation}
where
$f_{\lambda,1}(\beta):= \frac{1}{2}(\beta - \beta_1^*)^2 +
 2\lambda\abs{\beta}^{0.5}$
and
$f_{\lambda,2}(\beta):= \frac{1}{2}(\beta - \beta_2^*)^2 + 2\lambda\abs{\beta}^{0.5}$.
Figure \ref{fig:lambda_beta} plots the critical points
$\widetilde{\beta}_{1,\lambda}\in R_{\lambda}(\beta_1^*)\cup\{0\}$
and
$\widetilde{\beta}_{2,\lambda}\in R_{\lambda}(\beta_2^*)\cup\{0\}$
as a function of $\lambda$.
Recalling Example \ref{example:1D_critical_points},
$f_{\lambda,1}$ and $f_{\lambda,2}$ each have three critical points
within a certain range of $\lambda$,
and they each form three branches in Fig.~\ref{fig:lambda_beta}: A1, B1, and
 C1 for $f_{\lambda,1}$
and A2, B2, and C2 for $f_{\lambda,2}$.
Note that $f_{\lambda,1}(\beta_1)$ and $f_{\lambda,2}(\beta_2)$ are
coupled through a common $\lambda$.
Given a small $\lambda$, there are $9(=3\times 3)$
ways of choosing the pair of critical points
 $(\widetilde{\beta}_{1,\lambda},\widetilde{\beta}_{2,\lambda})$
from any pair of branches,
(A1,A2),
(A1,B2),
(A1,C2),
(B1,A2), $\cdots$,
(C1,C2).
Each pair forms a $\Qlt{0.5}$ path although
(C1,C2) is trivial as it corresponds to
the origin.
Excluding the trivial one,
there are eight other $\Qlt{0.5}$ paths 
(see Fig.~\ref{fig:c_trajectory}(a)).

For instance, let us start from $\lambda=0$ in Fig.~\ref{fig:lambda_beta}
and trace a critical path from the origin in Fig.~\ref{fig:c_trajectory}(a).
We increase $\lambda$ and follow the branches B1 and C2 until we reach
the edge of B1 at which A1 and B1 are connected.
This corresponds to the blue dotted line (labeled by B1C2)
in Fig.~\ref{fig:c_trajectory}(a).
Each point on the (B1,C2) path is a saddle point of $f_{\lambda}$,
since $\widetilde{\beta}_{1,\lambda}$ is a local maximum of $f_{\lambda,1}$
and $\widetilde{\beta}_{2,\lambda}=0$ is a local minimum of $f_{\lambda,2}$.
From the edge of B1,
we follow the branches A1 and C2 by decreasing $\lambda$ down to zero.
This corresponds to tracing the blue solid line (labeled by A1C2)
from the triangle.
Each point on the (A1,C2) path is a local minimum of $f_{\lambda}$
since both $\widetilde{\beta}_{1,\lambda}$ and 
$\widetilde{\beta}_{2,\lambda}=0$
are local minima of $f_{\lambda,1}$ and $f_{\lambda,2}$, respectively.
In an analogous way, one can associate every critical path
with a pair of branches in Fig.~\ref{fig:lambda_beta}.
Note that only the pair (B1,B2) gives local maxima of $f_{\lambda}$.

The union of the four paths (B1,C2), (A1,C2), (A1,B2), and (A1,A2)
of $\Qlt{0.5}$ forms a $\Pct{0.5}$ path
that starts from the origin and reaches the OLS solution
$\signal{\beta}^*$ through the breakpoint $[2,0]^{\sf T}$.
Fig.~\ref{fig:c_trajectory}(b) depicts each $\Pct{0.5}$ path in
 Fig.~\ref{fig:c_trajectory}(a) as a function of
$c$.
It can be seen that $c$ increases monotonically along any of the paths.
A question now is how $\lambda(c)$ changes with $c$ 
along the paths;
this is depicted in Fig.~\ref{fig:c_trajectory}(c).
It can be seen that $\lambda(c)$ is non-monotonic in $c$, and
the correspondence between $c$ and $\lambda(c)$ has multiplicity.
Note that the points marked by triangles along the paths in
Figs.~\ref{fig:c_trajectory}(a), (b) correspond to the peaks in
Fig.~\ref{fig:c_trajectory}(c) at which
$\dot{\lambda}(c):=\frac{d}{dc}\lambda(c)=0$
and a change from a local maximum to a local minimum in $\Ql{0.5}$ occurs.
Regarding the global solution paths,
the $\Pc{0.5}$ path is the whole blue curve in
 Fig.~\ref{fig:c_trajectory}(a), while
the $\Ql{0.5}$ path consists of
three disjoint sets $\{\signal{0}\}$, a subset of the (A1,C2) path,
and a subset of the (A1,A2) path (cf.~\cite{yuam_apsipa11}).
The parameter $c$ is a monotonically decreasing and {\it discontinuous}
function of $\lambda$.
\end{example}

In the non-orthogonal case, 
critical paths similar to the case of Example \ref{example:2D} are
obtained although the function $f_{\lambda}$ cannot be separated as in
\refeq{eq:f_lambda_separate}. 
(See Example \ref{example:2D_nonorthogonal} in the appendix.)

\subsection{Analysis}

Let us analyze critical paths of $\Pct{p}$,
while taking the critical point $\signal{\beta}(c)$ to be a function of $c\geq 0$.
How does $\signal{\beta}(c)$ behave
as $c$ changes?
The behavior can be described by
a differential equation governing the tangent direction 
$\dot{\signal{\beta}}(c):=\frac{d}{dc}\signal{\beta}(c)$
of the path $\signal{\beta}(c)$.
Let $\widetilde{\signal{\beta}}:=\signal{\beta}(c)$ and $\widetilde{\lambda}:=\lambda(c)$
in \refeq{eq:necessary_condition}, and let us
differentiate  both sides with respect to $c$.
After simple manipulations,
we obtain the equation of the critical path:
\begin{equation}
\signal{K}(\signal{\beta}(c))  \dot{\signal{\beta}}_{\mathcal{I}}(c)
= -\dot{\lambda}(c) 
\signal{\nabla}_{\mathcal{I}}
F_p(\signal{\beta}(c))
=\frac{\dot{\lambda}(c) }{\lambda(c)}
\signal{\nabla}_{\mathcal{I}}
\varphi(\signal{\beta}(c)).
\label{eq:path_characterization}
\end{equation}
One needs to carefully study those points at which the following situations occur.
\begin{enumerate}
 \item  The matrix $\signal{K}(\signal{\beta}(c))$
	is singular.
 \item $\dot{\lambda}(c)=0$.
 \item $\signal{\beta}(c)$ is a breakpoint where the support of $\signal{\beta}(c)$ changes.
\end{enumerate}

In Fig.~\ref{fig:c_trajectory}(a), 
the triangle indicates a separation point
$\signal{\beta}(c')$ of $\Qlt{0.5}$;
the smooth part of the path separates into a pair of $\Qlt{0.5}$ paths.
Viewing Fig.~\ref{fig:c_trajectory}(c), one can see that
$\dot{\lambda}(c')=0$ holds
at every separation point $\signal{\beta}(c')$ of $\Qlt{0.5}$ paths.
The matrix $\signal{K}(\signal{\beta}(c'))$ is also singular with
$\dot{\signal{\beta}}(c')$ being its eigenvector
associated with the zero eigenvalue,
since $\signal{\nabla}_{\mathcal{I}} F_p(\signal{\beta}(c'))$ is bounded
and $\dot{\signal{\beta}}_{\mathcal{I}}(c')\neq\signal{0}$.
The situations described in items 1) and 2) above happen
simultaneously, as shown by the following theorem.
\begin{theorem}[On singular points]
\label{theorem:singular}
On a $\Pct{p}$ path excluding its breakpoints and edges,
the following two statements are equivalent
if there is no other $\Pct{p}$ path passing through the point
$\signal{\beta}(c')$:
\begin{enumerate}
 \item [(a)] $\dot{\lambda}(c')=0$;
 \item [(b)] $\signal{K}(\signal{\beta}(c'))$ is singular.
\end{enumerate}
\end{theorem}
\noindent Proof: It has already been seen above that
(a) $\Rightarrow$ (b).
Assume that $\dot{\lambda}(c')\neq0$.
Suppose that $\signal{K}(\signal{\beta}(c'))$ is singular.
Then, since
$\signal{\nabla}_{\mathcal{I}}F_p(\signal{\beta}(c'))\neq \signal{0}$,
there is no $\dot{\signal{\beta}}_{\mathcal{I}}(c)$ satisfying
\refeq{eq:path_characterization}, or 
there are infinitely many $\dot{\signal{\beta}}_{\mathcal{I}}(c)$ satisfying
\refeq{eq:path_characterization} and the set of such
$\dot{\signal{\beta}}_{\mathcal{I}}(c)$s forms a linear variety which is unbounded.
This implies that the path is discontinuous.
Hence, $\signal{K}(\signal{\beta}(c'))$ should be nonsingular.
Indeed, the nonsingularity of $\signal{K}(\signal{\beta}(c'))$ ensures the existence
of a unique vector
$\dot{\signal{\beta}}_{\mathcal{I}}(c)$ that satisfies
\refeq{eq:path_characterization}.
This verifies that (b) $\Rightarrow$ (a).
\migip

Now consider a situation in which we follow
a critical path towards a breakpoint with $\beta_1>0$ approaching zero;
e.g., follow the (B1,~\!\!A2) path towards the breakpoint $[0,1]^{\sf T}$ in
Fig.~\ref{fig:c_trajectory}(a).
A simple inspection of \refeq{eq:necessary_condition} suggests that
$\widetilde{\lambda}=0$ and
$\partial_{i}\varphi(\widetilde{\signal{\beta}})=0$
for all $i\in\mathcal{I}\setminus\{1\}$ 
at the breakpoint, since
$\partial_{1}F_p(\widetilde{\signal{\beta}})\rightarrow \infty$ as 
$\widetilde{\beta}_1\uparrow 0$,
$\partial_{i}F_p(\widetilde{\signal{\beta}})< \infty$,
 $\forall i\in\mathcal{I}\setminus\{1\}$,
and $\partial_{i}\varphi(\widetilde{\signal{\beta}})< \infty$,
$\forall i\in\mathcal{I}$.
To analyze this situation in more detail, we will study the first component in
\refeq{eq:path_characterization}:
\begin{equation}
 \sum_{j\in\mathcal{I}} g_{1,j} \dot{\beta}_{j}(c)
-(1-p)\lambda(c)\beta_1^{-(2-p)}(c) \dot{\beta}_1(c)
= - \dot{\lambda}(c) \beta_1^{-(1-p)}(c),
\label{eq:first_comp}
\end{equation}
where it is assumed for simplicity that $\beta_1(c)> 0$.
Multiplying both sides of \refeq{eq:first_comp} by $\beta_1^{1-p}(c)$
and letting $\beta_1(c)\rightarrow 0$ yield
\begin{equation}
(1-p) \frac{\dot{\beta}_1(c)}{\beta_1(c)}
= \frac{\dot{\lambda}(c)}{\lambda(c)} 
~ \Leftrightarrow ~
(1-p) \frac{d}{dc}\log \beta_1(c) =
\frac{d}{dc}\log \lambda(c)
~ \Leftrightarrow ~
\beta_1^{1-p} (c)
\propto\lambda(c).
\label{eq:first_comp_lim}
\end{equation}
It is readily verified that
\begin{equation}
 \dot{\beta}_1(c)\propto \dot{\lambda}(c) \lambda^{\eta}(c)
\label{eq:beta1_dot}
\end{equation}
for $\eta:=\dfrac{p}{1-p}$.
Meanwhile, it holds that
\begin{equation}
 \partial_i \varphi(\signal{\beta})=
-\lambda(c) \partial_i F_p(\signal{\beta})=
-\lambda(c) {\rm sgn}(\beta_i)\abs{\beta_i^{-(1-p)}}, ~~~
\forall i\in\mathcal{I}.
\label{eq:partial_deriv}
\end{equation}
Let $\signal{\beta}_{\rm BR}$ denote a breakpoint
with its support $\mathcal{I}$
and $i'\in \bar{\mathcal{I}}$ an index that becomes active
at $\signal{\beta}_{\rm BR}$.
Then, we can verify the following theorem from
 \refeq{eq:first_comp_lim} -- \refeq{eq:partial_deriv}.
\begin{theorem}[Properties of breakpoints]
\label{theorem:breakpoint1}
At any breakpoint $\signal{\beta}_{\rm BR}=\signal{\beta}(c_{\rm BR})$,
it holds that
\begin{enumerate}
 \item $\lambda(c_{\rm BR})=0$;
 \item $\partial_i \varphi(\signal{\beta}_{\rm BR}) = 0, ~~~
 i\in\mathcal{I}$;\\
 $\partial_j \varphi(\signal{\beta}_{\rm BR}) \neq  0, ~~~ j\in
       \bar{\mathcal{I}}$.
\end{enumerate}
Moreover, $\dot{\beta}_{i'}(c)\rightarrow 0$
as $\signal{\beta}(c)$ with ${\rm supp}(\signal{\beta}(c))=\mathcal{I}\cup \{i'\}$
approaches the breakpoint $\signal{\beta}_{\rm BR}$.
\end{theorem}

Theorem \ref{theorem:breakpoint1}.1 (cf.~Fact \ref{fact:l1})
immediately yields the following corollary.

\begin{corollary}[Multiplicity and non-monotonicity of the $c$ -
 $\lambda$ correspondence for $0<p<1$]
\label{corollary:multiplicity}
Consider the correspondence between $c$ and $\lambda$ over
a path connecting the origin and an OLS solution.\footnote{
The path may not be a single critical path but could be composed of
a union of multiple critical paths.
}
Then, the following statements hold.
\begin{enumerate}
 \item $c(\lambda)$ is a multi-valued function of $\lambda\geq 0$.

 \item $\lambda(c)$ is a non-monotonic function of $c\geq 0$.
\end{enumerate}
\end{corollary}
\noindent Corollary \ref{corollary:multiplicity}.1
states that, given a $\lambda$ value,
there are multiple critical points $\signal{\beta}_{\lambda}$
that have different values of $c(\lambda):=F_p(\signal{\beta}_{\lambda})$.
From Theorem \ref{theorem:breakpoint1}.2, one can verify the following:
\begin{itemize}
 \item Every breakpoint is the best,
in the sense of minimizing $\varphi$,
among all points having the same support.\footnote{Some readers
may think that Theorem \ref{theorem:breakpoint1}.1 means
breakpoints can be obtained by solving $(\mathcal{L}_{0}^p)$.
This is, however, not true because 
the solution of $(\mathcal{L}_{0}^p)$ is clearly an OLS solution for any $p>0$.}
 \item  Any solution of $\Ql{p}$ for any $\lambda>0$ is {\it not} the best,
in the sense of minimizing $\varphi$, among all points
having the same support as the solution itself.
\end{itemize}





Finally, we  present the {\it connection theorem} at breakpoints.
Let $M$ denote the coordinate plane associated with
$\mathcal{I}\cup \{i'\}$ and $M_{\bar{i'}}(\subset M)$
the coordinate plane associated with $\mathcal{I}$.
On $M_{\bar{i'}}$, the critical-path equation
is given by
\begin{equation}
 \sum_{j\in\mathcal{I}} g_{i,j} \dot{\beta}_{j}(c)
-(1-p)\lambda(c)\beta_i^{-(2-p)}(c) \dot{\beta}_i(c)
= - \dot{\lambda}(c) \beta_i^{-(1-p)}(c),~~~i\in\mathcal{I},
\label{eq:ith_comp}
\end{equation}
which is identical to the critical-path equation on $M$
with $\dot{\beta}_{i'}(c)=0$.
This leads us to the following theorem.

\begin{theorem}[Connection theorem at breakpoints]
\label{theorem:connection}
Suppose that two smooth curves of critical points
are connected at a breakpoint.
Then, the curves touch tangentially at the breakpoint.
\end{theorem}

\section{Greedy Path and Its Link to OMP}
\label{sec:Pc_OMP}

In this section, we consider two continuous paths of critical points,
{\it a main path} and {\it a greedy path}, in the overdetermined case.

\subsection{Main Path and Greedy Path}

The main path is a continuous curve from the OLS solution
$\signal{\beta}^*$ to the origin;
e.g., the blue curves
in Figs.~\ref{fig:c_trajectory}(a) and \ref{fig:critical_paths},
and the union of the green, red, and blue curves in
Fig.~\ref{fig:main_path}(a) (see the appendix).
To be precise, the main path is defined as follows.
\begin{definition}[Main path]
~
\begin{enumerate}
 \item A main path starts from the $\signal{\beta}^*$
(the initial active-index set is $\mathcal{I}_0:=\{1,2,\cdots,n\}$ generically)
and follows the critical-path equation \refeq{eq:path_characterization}.
 \item If it reaches a breakpoint where some variable, say
       $\beta_{i^{\star}}$, becomes zero,
then the path follows \refeq{eq:path_characterization}
with the  updated active-index set
$\mathcal{I}_1:=\mathcal{I}_0\setminus\{i^{\star}\}$.
 \item If it reaches the next breakpoint where another variable, say
       $\beta_{j^{\star}}$, becomes zero,
then the path follows \refeq{eq:path_characterization}
with 
$\mathcal{I}_2:=\mathcal{I}_1\setminus\{j^{\star}\}=\mathcal{I}_0\setminus\{i^{\star},j^{\star}\}$.
\item Repeat the same procedure until the path reaches the origin.
\end{enumerate}
\end{definition}
On the other hand, a greedy path is a continuous curve which starts at the origin 
and possibly ends at $\signal{\beta}^*$.
It is an extension of the LARS path to the case of $p<1$ and
provides a remarkable link between
the $\ell_p$-regularized least squares and OMP.
The greedy path is defined as follows.
\begin{definition}[Greedy path]
~
\begin{enumerate}
 \item A greedy path starts from the origin
and follows the critical-path equation \refeq{eq:path_characterization}
with $\mathcal{I}_0:=\{i^{\star}\}$
for $i^{\star}\in\argmax_{i=1,2,\cdots,n}
       \abs{\partial_i\varphi(\signal{0})}$.
At the origin, \refeq{eq:path_characterization} suggests the direction\footnote{
This is because (i) $\lambda(c) \geq 0$, 
(ii) $\dot{\lambda}(c)>0$ in the vicinity of the origin since
       $\lambda(c)\rightarrow 0$ as $\signal{\beta}\rightarrow\signal{0}$, and
(iii) $\signal{K}(\signal{\beta})\rightarrow -\infty$ as $\signal{\beta}$
approaches $\signal{0}$ along some coordinate.
Indeed, as $\signal{\beta}$
approaches $\signal{0}$ along some coordinate,
$\dot{\lambda}(c)\rightarrow \kappa\in(0,\infty)$
and $\signal{K}(\signal{\beta})\lambda(c)\rightarrow -\infty$
so that $\dot{\beta}_{i'}(c)\rightarrow 0$,
$i'\in{\rm supp}(\signal{\beta})$
(see Theorem \ref{theorem:breakpoint1}.1).
}
$[0,\cdots,0,-\partial_{i^{\star}}\varphi(\signal{0}),0,
\cdots,0]^{\sf T}$.

 \item Once it reaches a breakpoint $\signal{\beta}_{\rm BR}^{1}$ where
       $\partial_{i^{\star}}\varphi(\signal{\beta}_{\rm BR}^{1})=0$, 
the path follows \refeq{eq:path_characterization}
with the updated active-index set
$\mathcal{I}_1:=\mathcal{I}_0\cup\{j^{\star}\}=\{i^{\star},j^{\star}\}$
for $j^{\star}\in\argmax_{j=1,2,\cdots,n} \abs{\partial_j\varphi(\signal{\beta}_{\rm BR}^{1})}$.

 \item Once it reaches the next breakpoint $\signal{\beta}_{\rm BR}^{2}$ where
       $\partial_{i^{\star}}\varphi(\signal{\beta}_{\rm BR}^{2})=\partial_{j^{\star}}
\varphi(\signal{\beta}_{\rm BR}^{2})=0$,
the path follows \refeq{eq:path_characterization}
with 
$\mathcal{I}_2:=\mathcal{I}_1\cup\{k^{\star}\}=\{i^{\star},j^{\star},k^{\star}\}$
for $k^{\star}\in\argmax_{k=1,2,\cdots,n}
       \abs{\partial_k\varphi(\signal{\beta}_{\rm BR}^{2})}$.

\item Repeat the same procedure until the path reaches
      $\signal{\beta}^*$.
(The path would stop if some variable became zero accidentally.)

\end{enumerate}
\end{definition}

Suppose, in the first step of the greedy path,
 that \refeq{eq:path_characterization} suggests 
an undesirable direction in the sense that
the path leads to the opposite side
from $\signal{\beta}^*$ with respect to its $i^{\star}$th
component.
Such an $i^{\star}$ could be excluded from the active-index selection, since
the path cannot reach $\signal{\beta}^*$
without getting $\beta_{i^{\star}}$ back to zero.
We thus define the modified greedy path as follows.

\begin{definition}[Modified Greedy Path]
\label{def:modification}
In the first step to finding the greedy path,
let $i^{\star}\in\argmax_{i\in\mathcal{J}_0}
       \abs{\partial_i\varphi(\signal{0})}$,
where $\mathcal{J}_0:=\{i=1,2,\cdots,n:\partial_i
 \varphi(\signal{0})\beta_i^*<0 \}$.
In the second step,
let $j^{\star}\in\argmax_{i\in\mathcal{J}_1}
       \abs{\partial_i\varphi(\signal{\beta}_{\rm BR}^{1})}$,
where $\mathcal{J}_1:=\{i=1,2,\cdots,n:\partial_i
 \varphi(\signal{\beta}_{\rm BR}^{1})
\beta_i^*<0 \}$.
The same applies to the subsequent steps.
\end{definition}

For a fixed $\mathcal{I}$, both the main and greedy paths are smooth because
their directions are
governed by \refeq{eq:path_characterization}.
The way of selecting active indices in the greedy path will be validated
in Section \ref{subsec:minkowski} by using a generalized Minkowskian gradient.
In Examples 
\ref{example:1D_critical_points}, 
\ref{example:2D}, 
\ref{example:2D_nonorthogonal}, 
\ref{example:p07},
the greedy paths coincide with the main paths, and those in Examples
\ref{example:2D}, 
\ref{example:2D_nonorthogonal}, 
\ref{example:p07}
 are homeomorphic with each other.
Note that $c$ is not necessarily monotonic along the main/greedy path
(see Example \ref{example:p07}).
A particular case in which 
the modification is required for the greedy path
is Example \ref{example:counter_example_3D} in the appendix.

Important observations regarding the relation between 
the four paths (global solution path, critical paths,
main path, and greedy path) are summarized below.

\begin{observation}
\label{observation:greedy_path}
 ~
\begin{enumerate}
 \item Generically, there is a unique main path
and a unique greedy path.\footnote{In an exceptional case, for instance,
       in which
       $\signal{\beta}^*:=[1,1]^{\sf T}$, $\signal{G}:=\signal{I}$,
$p=0.5$, the main path starts at $\signal{\beta}^*$ in the direction
towards the origin
and splits into three paths: one goes to the origin straightly and 
the others respectively go to the origin via the breakpoints $[0,1]^{\sf T}$
and $[1,0]^{\sf T}$ due to symmetry.

}
 \item The main path, greedy path, and global solution path
are subsets of $\mathcal{C}$.


 \item The main path (the greedy path) is composed of a union of multiple $\Qlt{p}$ paths.
 \item The main path (the greedy path) is either a single $\Pct{p}$ path
       or  composed of a
       union of multiple $\Pct{p}$ paths (see Example \ref{example:p07}
       in the appendix).
 \item When $\signal{G}=\signal{I}$, 
the main and global solution paths coincide with each other, or otherwise
the main path includes the global
       solution path as its subset (see Example \ref{example:p07}).
\end{enumerate}

\end{observation}

\begin{remark}[Underdetermined case]
In the underdetermined case, there are infinitely many OLS solutions.
The main path can still be defined as the one starting
from a sparsest OLS solution $\signal{\beta}^*$.
In this case, however, it is not useful for solving a sparse
optimization problem because its starting point is a solution of the problem.
The greedy path is, however, useful.
The minimum-norm OLS solution 
$\signal{\beta}^*:=(\signal{X}^{\sf T})^{\dagger}\signal{y}$
can be used to determine the modification process.
\end{remark}

\subsection{Generalized Minkowskian Gradient and Greedy Path}
\label{subsec:minkowski}

We show that part of the greedy path can be constructed with
a generalized Minkowskian gradient.
See \cite{amari13} for a study of the Minkowskian gradient for sparse optimization
with $p=1$, which encompasses non-quadratic convex objective-functions.
To define a generalized Minkowskian gradient,
we introduce a pseudo-norm below.
\begin{definition}
Given any vector $\signal{\beta}\in\real^n$ 
with ${\rm supp}(\signal{\beta})=\mathcal{I}$
such that the Hessian matrix
$\signal{K}(\signal{\beta})$ is positive definite, 
 we define the pseudo-norm of a vector $\signal{a}\in\real^n$, depending
 on the position $\signal{\beta}$, by
\begin{equation}
 Q_{\signal{\beta}}(\signal{a}):=
\sqrt{\signal{a}_{\mathcal{I}}^{\sf T}
\signal{K}(\signal{\beta})\signal{a}_{\mathcal{I}}}
+ \frac{1}{p} \sum_{i\in \bar{\mathcal{I}}} \psi_p(a_i).
\end{equation}
\end{definition}

\begin{definition}
Given any vector $\signal{\beta}\in\real^n$ such that
$\signal{K}(\signal{\beta})$ is positive definite, 
the generalized Minkowskian gradient of $\varphi(\signal{\beta})$
is defined as follows:
\begin{equation}
 \signal{\nabla}_{\rm GM}\varphi(\signal{\beta}):=
\argmax_{Q_{\signal{\beta}}(\signal{a})=1} 
\signal{a}^{\sf T}\signal{\nabla}\varphi(\signal{\beta}).
\end{equation}
\end{definition}

\begin{lemma}[Generalized Minkowskian gradient at  $\signal{\beta}=\signal{0}$]
\label{lemma:gm1}
The generalized Minkowskian gradient
at the origin is given by
\begin{equation}
 [\signal{\nabla}_{\rm GM}\varphi(\signal{0})]_i= 
\left\{
\begin{array}{ll}
-{\rm sgn}(\partial_{i^{\star}}\varphi(\signal{0})),  & ~~~i^{\star}\in
 \displaystyle\argmax_{\iota=1,2,\cdots,n} \abs{\partial_{\iota}
\varphi(\signal{0})}, \\
0, & ~~~ i\neq i^{\star},
\end{array}
\right.~~~\forall i=1,2,\cdots,n.
\label{eq:gm_1}
\end{equation}
\end{lemma}
Proof: The pseudo-norm $Q_{\signal{0}}$ coincides with the $\ell_p$
quasi-norm, and the generalized Minkowskian gradient is equivalent to
the Minkowskian gradient for $p=1$, as stated in \refeq{eq:gm_1}.
This can readily be verified
by the concavity of the $\ell_p$ quasi-norm in each orthant.
\migip

\begin{lemma}[Generalized Minkowskian gradient 
at $\signal{\beta}$,
 $\beta_i\neq 0$, $\forall i=1,2,\cdots,n$]
\label{lemma:gm2}
For $\signal{\beta}$ with
 $\beta_i\neq 0$, $\forall i=1,2,\cdots,n$,
the generalized Minkowskian gradient is given by
\begin{equation}
\signal{\nabla}_{\rm GM}\varphi(\signal{\beta})=
\frac{\signal{K}^{-1}(\signal{\beta})\signal{\nabla}\varphi(\signal{\beta})}
{\sqrt{\signal{\nabla}\varphi(\signal{\beta})^{\sf T}
\signal{K}^{-1}(\signal{\beta})\signal{\nabla}\varphi(\signal{\beta})}}
\propto \signal{K}^{-1}(\signal{\beta})\signal{\nabla}\varphi(\signal{\beta}).
\label{eq:gm_2}
\end{equation}
\end{lemma}
Proof: The claim is readily verified with a Lagrange multiplier.

\begin{lemma}[Generalized Minkowskian gradient at a general $\signal{\beta}$]
\label{lemma:gm3}
~
\begin{enumerate}
 \item 
Let $\signal{\nabla}_{\mathcal{I}} \varphi(\signal{\beta})\neq \signal{0}$.
Then,
\begin{align}
\signal{\nabla}_{{\rm GM},\mathcal{I}} \varphi(\signal{\beta})=& ~
\frac{\signal{K}^{-1}(\signal{\beta})\signal{\nabla}_{\mathcal{I}}\varphi(\signal{\beta})}
{\sqrt{\signal{\nabla}_{\mathcal{I}} \varphi(\signal{\beta})^{\sf T}
\signal{K}^{-1}(\signal{\beta})
 \signal{\nabla}_{\mathcal{I}}\varphi(\signal{\beta})}}
\propto
 \signal{K}^{-1}(\signal{\beta})\signal{\nabla}_{\mathcal{I}}\varphi(\signal{\beta}),\\
[\signal{\nabla}_{{\rm GM}} \varphi(\signal{\beta})]_i =& ~
0,~~~i\in\bar{\mathcal{I}}.
\label{eq:gm_3a}
\end{align}
\item 
Let $\signal{\nabla}_{\mathcal{I}} \varphi(\signal{\beta}) =
 \signal{0}$; i.e., let $\signal{\beta}$ be a breakpoint.
Then,
\begin{align}
\signal{\nabla}_{{\rm GM},\mathcal{I}} \varphi(\signal{\beta})=& ~
\signal{0},\\
 [\signal{\nabla}_{\rm GM}\varphi(\signal{\beta})]_i= & ~
\left\{
\begin{array}{ll}
-{\rm sgn}(\partial_{i^{\star}}\varphi(\signal{\beta})),  &
 ~~~i^{\star}\in 
\displaystyle\argmax_{\iota=1,2,\cdots,n} \abs{\partial_{\iota}
\varphi(\signal{\beta})}, \\
0, & ~~~ i\neq i^{\star},
\end{array}
\right.~~~\forall i\in\bar{\mathcal{I}}.
\label{eq:gm_3b}
\end{align}
\end{enumerate}
\end{lemma}
Proof: The pseudo-norm
$Q_{\signal{\beta}}(\signal{a})$
is a first-order function of $a_i$ for $i\in\mathcal{I}$ while
it is a $p$th order function of $a_i$ for $i\in\bar{\mathcal{I}}$.
Since $p<1$, in order to maximize
$\signal{a}^{\sf T}\signal{\nabla}\varphi(\signal{\beta})
=\signal{a}_{\mathcal{I}}^{\sf T}
\signal{\nabla}_{\mathcal{I}}\varphi(\signal{\beta})
+\signal{a}_{\bar{\mathcal{I}}}^{\sf T}
\signal{\nabla}_{\bar{\mathcal{I}}}\varphi(\signal{\beta})$,
all resources should be allocated to $\signal{a}_{\mathcal{I}}$,
if $\signal{\nabla}_{\mathcal{I}}\varphi(\signal{\beta})\neq
\signal{0}$, and to $\signal{a}_{\bar{\mathcal{I}}}$,
if $\signal{\nabla}_{\mathcal{I}}\varphi(\signal{\beta}) = \signal{0}$.
This verifies the claim.
\migip

Lemmas \ref{lemma:gm1}--\ref{lemma:gm3} lead to the following theorem.
\begin{theorem}
 \label{theorem:generalized_minkowski}
The direction vector of the greedy path
is given by 
$-\signal{\nabla}_{{\rm GM}} \varphi(\widetilde{\signal{\beta}})$
at any point $\widetilde{\signal{\beta}}$ where
$\signal{K}(\widetilde{\signal{\beta}})$ is positive definite, including
 the origin and all the breakpoints.
\end{theorem}
Note here that $\dot{\lambda}(c)<0$ in \refeq{eq:path_characterization}
when $\signal{K}(\widetilde{\signal{\beta}})$ is positive definite
(cf.~Theorem \ref{theorem:singular}).
Note also that,
when $\signal{K}(\widetilde{\signal{\beta}})$ has a negative eigenvalue,
the direction vector of the greedy path 
on the coordinate plane associated with the active-index set $\mathcal{I}$
is given by
$\signal{K}^{-1}(\widetilde{\signal{\beta}})\signal{\nabla}_{\mathcal{I}}\varphi(\widetilde{\signal{\beta}})$,
rather than
$-\signal{K}^{-1}(\widetilde{\signal{\beta}})\signal{\nabla}_{\mathcal{I}}\varphi(\widetilde{\signal{\beta}})$.
This is because the direction vector 
in this case
is $\dot{\signal{\beta}}(c)$ and
$\dot{\lambda}(c)>0$ if $c$ increases along the greedy path,
while 
the direction vector is $-\dot{\signal{\beta}}(c)$ and
$\dot{\lambda}(c)<0$ if $c$ decreases.
Special care is therefore required at those points where
the Hessian matrix $\signal{K}(\widetilde{\signal{\beta}})$ is singular.

\subsection{Link Between $\ell_p$-Regularized Least Squares and OMP}

The following proposition immediately follows from the
definition of the greedy path.
\begin{theorem}
[Link between OMP and the $\ell_p$ regularized least squares]
\label{theorem:link_Pc}
Suppose that the (unmodified) greedy path continues
to an OLS solution.
Then, the breakpoints of the greedy path coincide with
the step-by-step solutions generated by OMP.
\end{theorem}

\begin{corollary}
[Link between OMP and $\Ql{p}$]
\label{corollary:link_Pl}
Suppose that the (unmodified) greedy path continues
to an OLS solution.
 Each step-by-step solution generated by OMP is
the limit of a convergent sequence of critical points of $\Ql{p}$ as
 $\lambda\rightarrow 0$.
\end{corollary}
\noindent Proof: 
The claim is readily verified
using Theorems \ref{theorem:breakpoint1}.1 and \ref{theorem:link_Pc}.
\migip

The link between OMP and the $\ell_p$-regularized
least squares presented in Theorem \ref{theorem:link_Pc}
is more direct than the one between OMP and $\ell_1$ minimization.
Theorem \ref{theorem:link_Pc} 
naturally leads us to the modified OMP algorithm below.

\begin{algorithm}[Modified OMP Algorithm]
\label{alg:modified_omp}
Compute the breakpoints of the modified greedy path
one by one in the same way as OMP;
i.e., minimize $\varphi$, at each step, in terms of active variables
with inactive variables being zero.
\end{algorithm}

Let us have a fresh look at Example \ref{example:2D}.
It is easily verified that 
the step-by-step solutions of OMP in the example are
$\signal{\beta}_1^{\rm OMP}:=[2,0]^{\sf T}$ and
$\signal{\beta}_2^{\rm OMP}:=[2,1]^{\sf T}$.
One can see that 
$\signal{\beta}_1^{\rm OMP}$ is the breakpoint
of the greedy path B1C2 -- A1C2 -- A1B2 -- A1A2 (the blue curve in Fig.~\ref{fig:c_trajectory}(a)),
and $\signal{\beta}_2^{\rm OMP}$
is the end point of A1A2, which is the OLS solution.
This clearly demonstrates Theorem \ref{theorem:link_Pc}.


\section{Conclusion}
\label{sec:conclusion}

This paper investigated
the least squares problem by making two different formulations
involving $\ell_p$-regularization ($0<p<1$):
 $\ell_p$-constrained least squares $\Pc{p}$ and
 $\ell_p$-penalized least squares $\Ql{p}$.
The key findings are summarized as follows.
\begin{enumerate}
 \item The essential difference between $\Pc{p}$ and $\Ql{p}$:
the $\Ql{p}$-path is a proper subset of the $\Pc{p}$-path
(Theorem \ref{theorem:proper_subset}).
The two problems are also different in terms of their local minimality
       (Lemma \ref{lemma:ns}).

 \item Discontinuity of the solution paths:
the $\Ql{p}$ solution paths are always discontinuous,
whereas the $\Pc{p}$ solution paths are possibly discontinuous
(Lemma \ref{lemma:continuity} and Example \ref{example:p07}).
This is due to the nonconvexity of the $\ell_p$ quasi-norm.

 \item Properties of breakpoints: $\lambda(c)=0$
at any breakpoint
(Theorem \ref{theorem:breakpoint1}.1).
Moreover, every breakpoint is the best, in the sense of minimizing $\varphi$,
among all points having the same support 
(Theorem \ref{theorem:breakpoint1}.2).
Two smooth curves connected at a breakpoint
touch tangentially (Theorem \ref{theorem:connection}).

 \item Multiplicity (non-monotonicity) in the correspondence between the
       regularization parameters: 
multiple $c$ values in $\Pc{p}$ correspond to a
       single value of $\lambda$ in $\Ql{p}$ (Corollary
       \ref{corollary:multiplicity}).
 \item Greedy path and generalized Minkowskian gradient:
the direction vector of the greedy path
is given by the generalized Minkowskian gradient
at any point where the Hessian matrix $\signal{K}(\signal{\beta})$ 
is positive definite (Theorem \ref{theorem:generalized_minkowski}).

 \item The direct link between OMP and $\ell_p$-regularized least
       squares:
the breakpoints of the greedy path
coincide with OMP step-by-step solutions
(Theorem \ref{theorem:link_Pc}).
The link is more direct than that between OMP and $\ell_1$ minimization.
\end{enumerate}

It should be remarked that some parts of the greedy path are not covered
by the theory presented in \cite{chen10,xu12}.
Indeed, what is obtained by
the existing approximate solvers for $\Ql{p}$ given some $\lambda>0$
is a stable critical point of $\Ql{p}$, which is not 
necessarily on the greedy path.
The fundamental study on critical paths presented here will be
a useful basis
for making the output of an $\ell_p$-regularization-based approach
more controllable.
Developing a computational method to construct a main/greedy path
will be an interesting future work.

\appendices


\newcounter{appnum}
\setcounter{appnum}{1}

\setcounter{theorem}{0}
\renewcommand{\thetheorem}{\Alph{appnum}.\arabic{theorem}}

\setcounter{lemma}{0}
\renewcommand{\thelemma}{\Alph{appnum}.\arabic{lemma}}
\setcounter{example}{0}
\renewcommand{\theexample}{\Alph{appnum}.\arabic{example}}
\setcounter{equation}{0}
\renewcommand{\theequation}{\Alph{appnum}.\arabic{equation}}

\section{Examples}
\label{append:examples}

This appendix presents four examples:
\begin{enumerate}
 \item [{\bf A.1}] Non-orthogonal case ($\signal{G}\neq \signal{I}$) for
       $n=2$ and $p=0.5$.
This is a simple example of critical paths for a non-orthogonal case.

 \item [{\bf A.2}] Orthogonal case ($\signal{G} = \signal{I}$) for
       $n=2$ and $p=0.7$.
This is a particular case in which (i)
the $\Pc{p}$ solution path is discontinuous, and 
(ii) $c$ is non-monotonic along the main/greedy path.

 \item [{\bf A.3}] Non-orthogonal case for $n=3$ and $p=0.5$.
This is a particular case in which
a modification must be made to get the greedy path
(see Definition \ref{def:modification}).
 \item [{\bf A.4}] Orthogonal case for $n=5$ and $p=0.5$.
This is an example of greedy paths for a higher dimensional system.

\end{enumerate}

\begin{figure}[t]
\begin{center}
\psfrag{b1}[c][1.0]{$\beta_1$}
\psfrag{b2}[c][1.0]{$\beta_2$}
 \includegraphics[width=8cm]{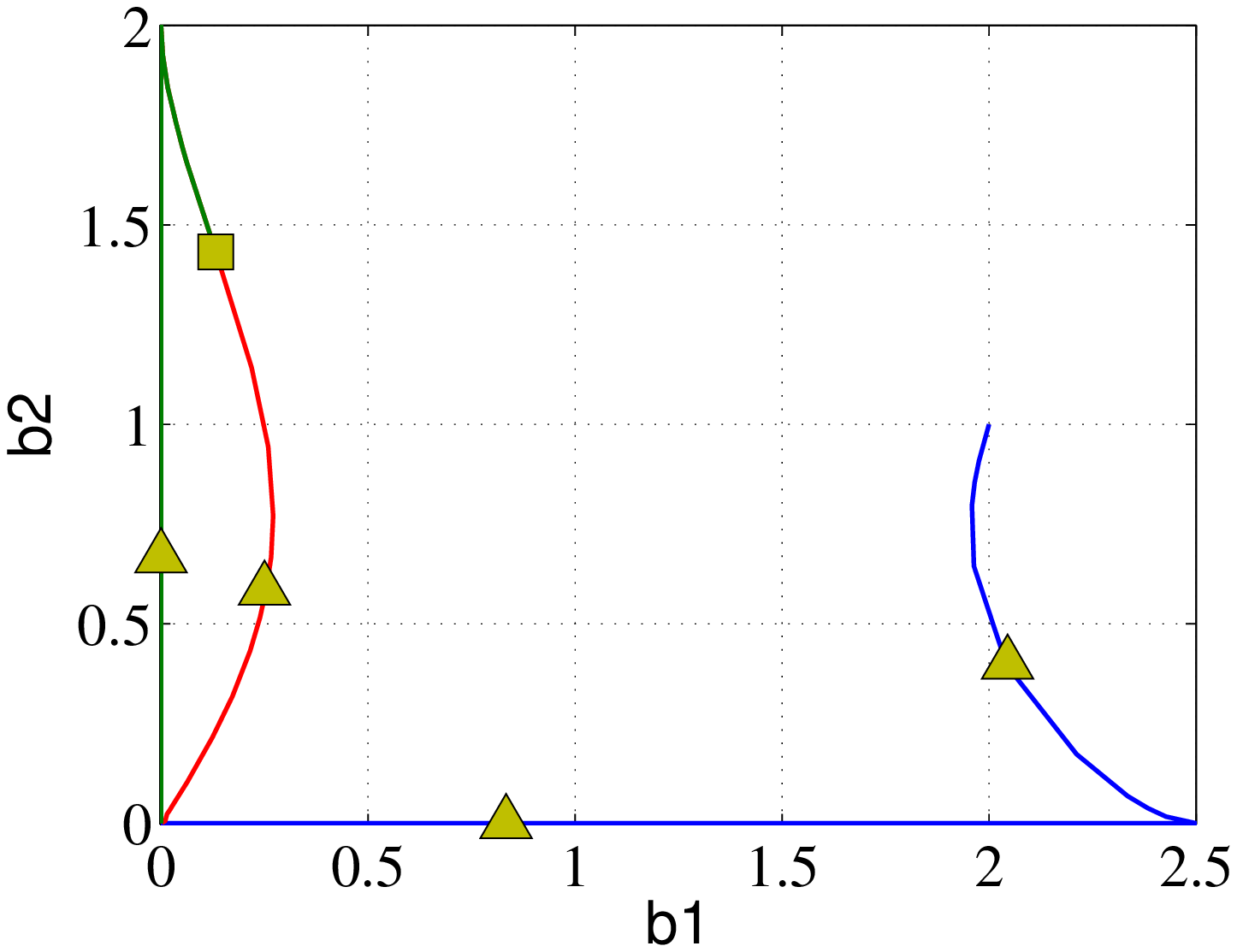} 
\end{center}
\caption{Critical paths of $\Pct{0.5}$ in Example
 \ref{example:2D_nonorthogonal}.}
\label{fig:critical_paths}
\end{figure}

\begin{example}[2D non-orthogonal case]
\label{example:2D_nonorthogonal}
Consider the following example:
 $\varphi(\signal{\beta}):=\frac{1}{2}
\norm{\signal{\beta} - \signal{\beta}^*}_{\signal{G}}^2:=
\frac{1}{2}
(\signal{\beta} - \signal{\beta}^*)^{\sf T}\signal{G}
(\signal{\beta} - \signal{\beta}^*)$ 
with $\signal{\beta}^*:=[2,1]^{\sf T}$,
$\signal{G}:=\left[\begin{array}{cc}
	      1&0.5 \\
	      0.5 &1 \\
		   \end{array}\right]$, and
$F_{0.5}(\signal{\beta}):=2(|\beta_1|^{0.5} + |\beta_2|^{0.5})$.
In this case, there are three $\Pc{0.5}$ paths;
Fig.~\ref{fig:critical_paths} shows
the critical paths drawn in different colors.
Unlike the case of $\signal{G}=\signal{I}$ in Example \ref{example:2D},
the function $f_{\lambda}$ cannot be separated as in
\refeq{eq:f_lambda_separate} and, therefore, one should consider 
both variables $\beta_1$ and $\beta_2$ together in order to find the critical points.
In the general case of $n\geq 2$,
the partial derivatives $\partial_i f_{\lambda}(\signal{\beta})$ for
 $i\in{\rm supp}(\signal{\beta})$
depend on the other variables, and the condition for $\signal{\beta}$
 to be a critical point is given by
\begin{equation}
\beta_i  + \alpha_i + \lambda \abs{\beta_i}^{p-1}{\rm sgn}(\beta_i) = 0,
~~~\forall i\in{\rm supp}(\signal{\beta}),
\label{eq:cond_equivalent_simple}
\end{equation}
where
\begin{equation}
 \alpha_i:=\alpha_i(\beta_1,\beta_2,\cdots,\beta_{i-1},\beta_{i+1},\cdots,\beta_n):=
- \beta_i^*  + \sum_{j\neq i} g_{i,j} (\beta_j  - \beta_j^* ).
\end{equation}
Here, $g_{i,j}$ is the $(i,j)$ component of $\signal{G}$.
\end{example}

\begin{figure}[t]
\centering
\begin{psfrags}
\psfrag{c}[c][1.0]{$c$}
\psfrag{beta1}[c][1.0]{$\beta_1$}
\psfrag{beta2}[c][1.0]{$\beta_2$}
\subfigure[]{
 \includegraphics[width=10cm]{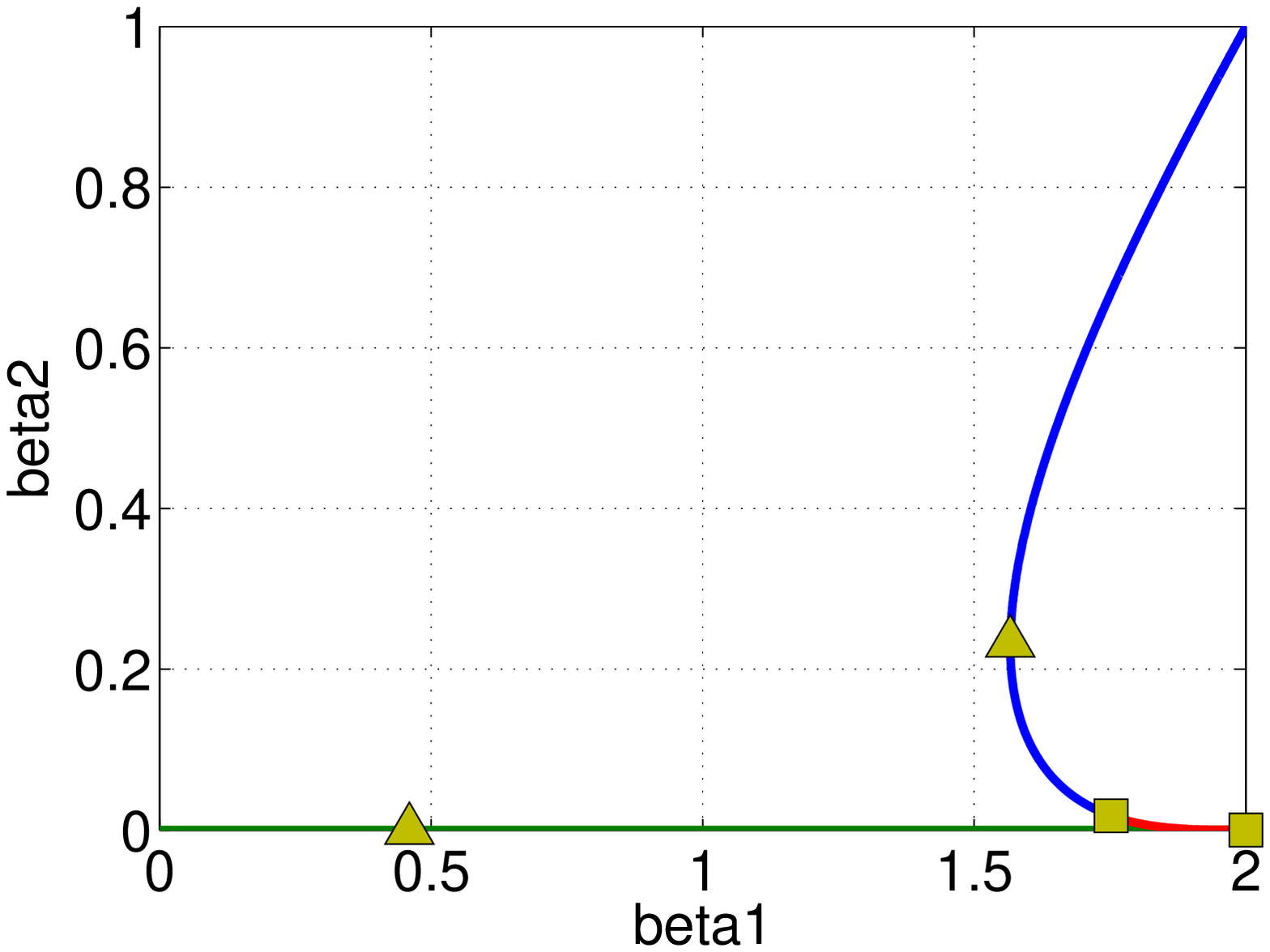}
}
\subfigure[]{
 \includegraphics[width=10cm]{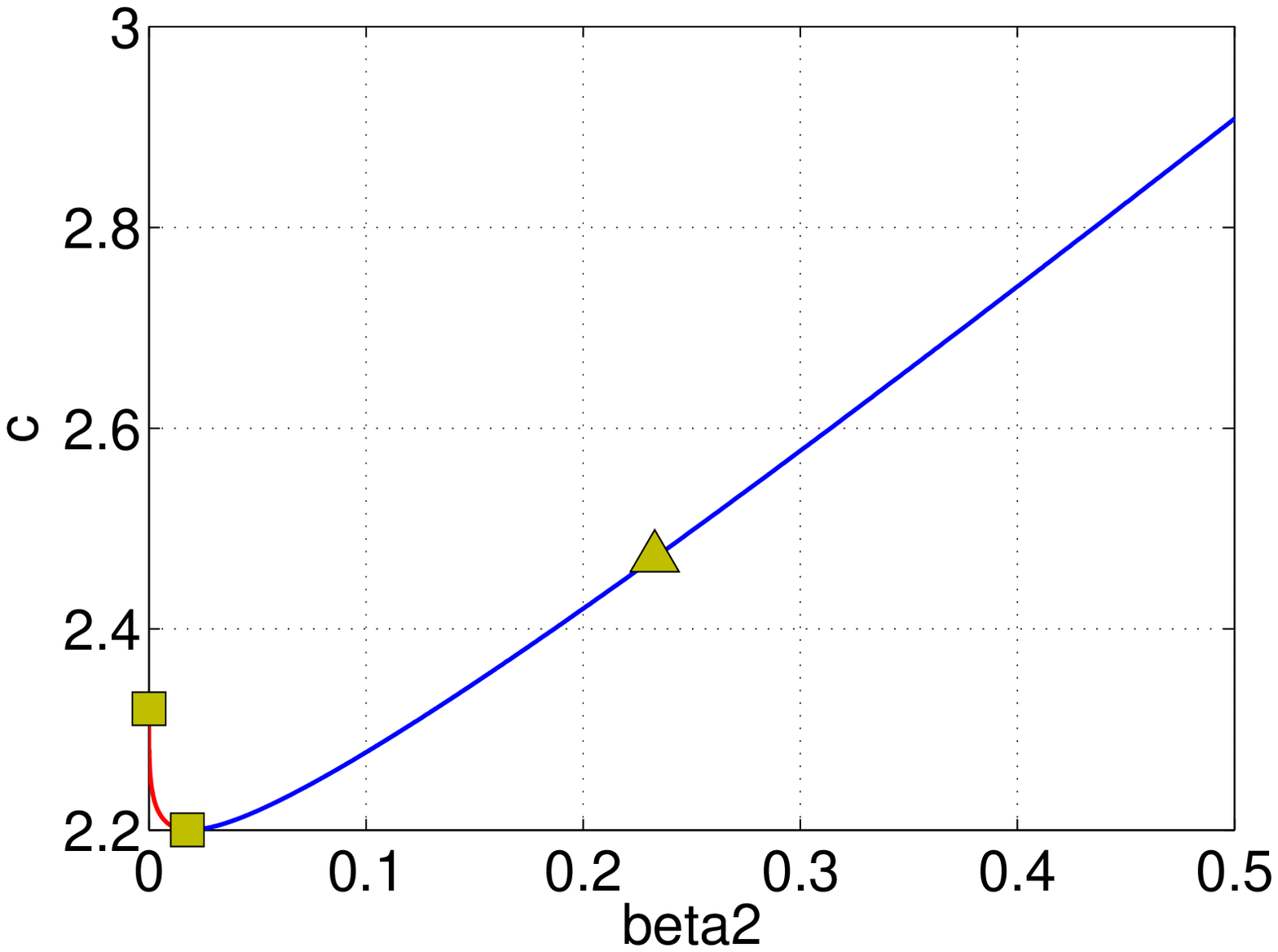}
}
\end{psfrags}
 \caption{(a) The main path composed of a union of 
three $\Pct{0.7}$ paths and (b)
 non-monotonicity of $c$ 
along the main path in Example \ref{example:p07}.}
\vspace*{0em}
\label{fig:main_path}
\end{figure}

\begin{figure}[t]
\centering
\begin{psfrags}
\psfrag{c}[c][1.0]{$c(=F_p(\signal{\beta}))$}
\psfrag{phi}[c][1.0]{$\varphi(\signal{\beta})$}
\psfrag{beta1}[c][1.0]{$\beta_1$-coordinate}
\psfrag{beta12}[c][1.0]{$\beta_1$-$\beta_2$ plane}
\subfigure[]{
 \includegraphics[width=10cm]{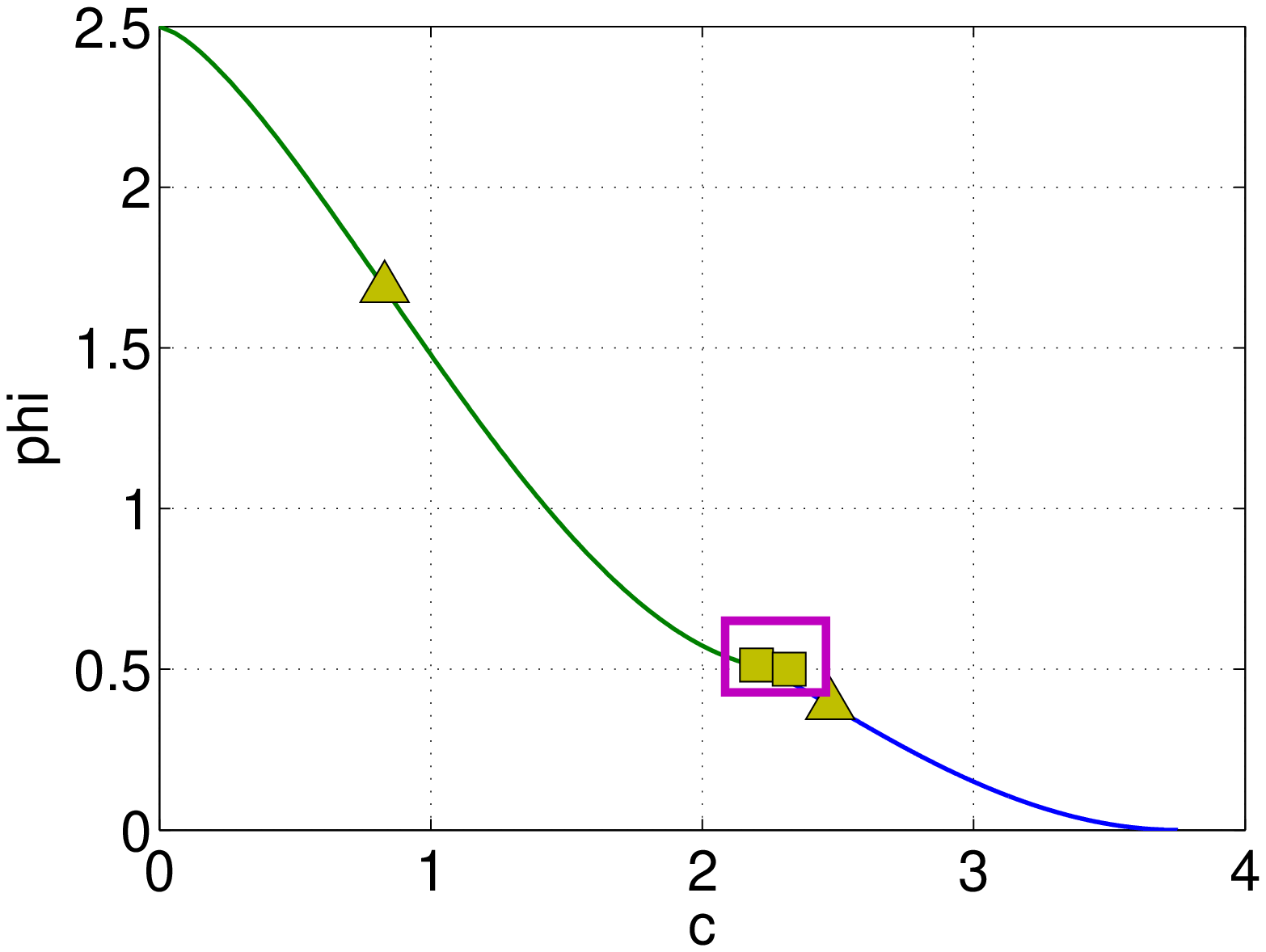}
}
\subfigure[]{
 \includegraphics[width=10cm]{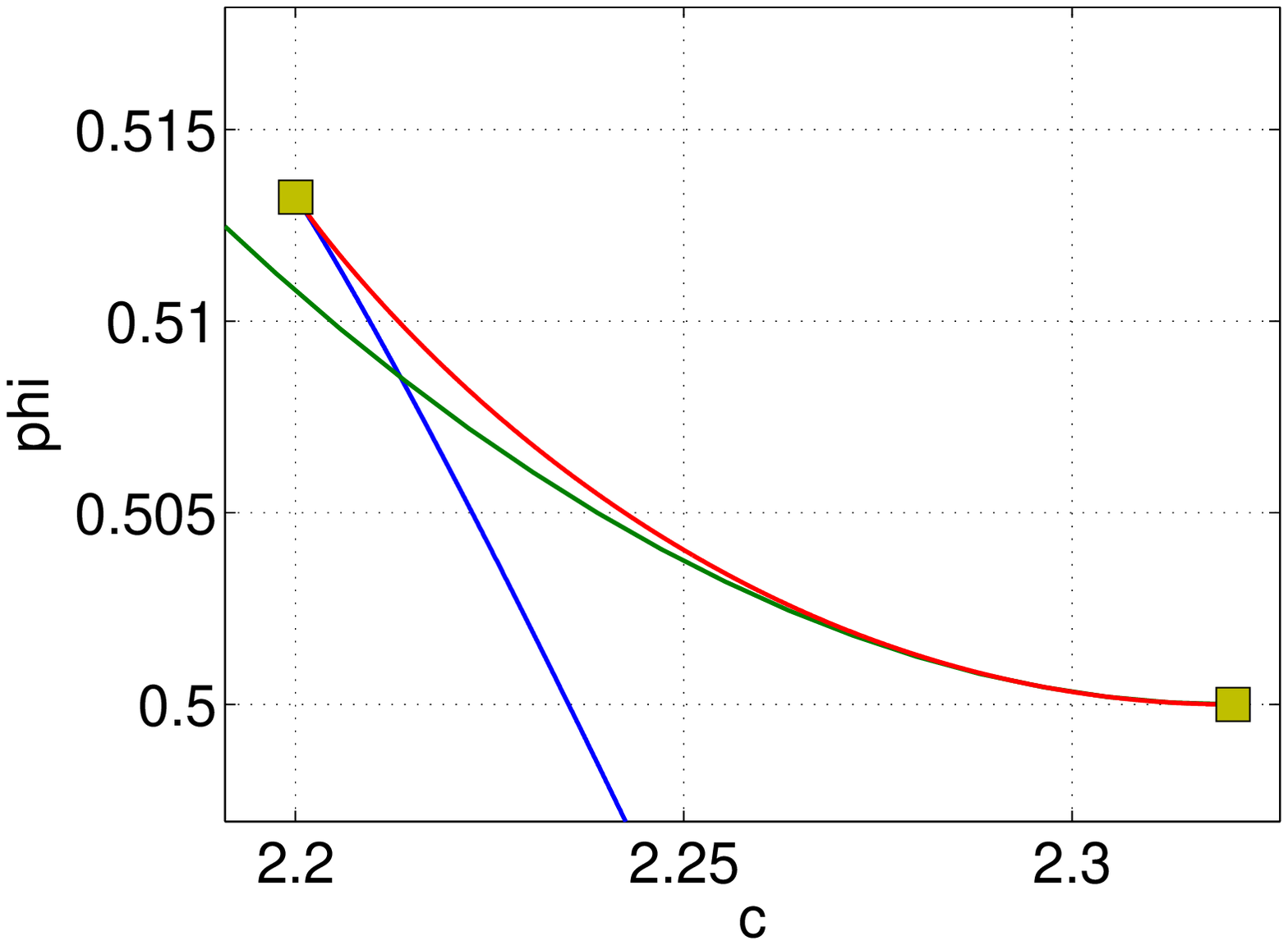}
}
\end{psfrags}
\caption{(a) $\varphi$ values along the main path
in Example \ref{example:p07} and
(b) zooming in on the purple square in (a).}
\vspace*{0em}
\label{fig:c_phi}
\end{figure}

\begin{example}[2D orthogonal case for $p=0.7$]
\label{example:p07}
 Consider the following case of $n=2$:
 $\varphi(\signal{\beta}):=\frac{1}{2}
\norm{\signal{\beta} - \signal{\beta}^*}_2^2$ with
$\signal{\beta}^*:=[2,1]^{\sf T}$
and
$F_{p}(\signal{\beta}):=\frac{1}{p}(|\beta_1|^p + |\beta_2|^p )$ for $p=0.7$.
Although the only difference from Example \ref{example:2D} is the $p$
value, it leads to significant differences as explained below.
\begin{enumerate}
 \item $c$ is non-monotonic along the path connecting the origin and
       $\signal{\beta}^*$ (which will be referred to as the main path in
       Section \ref{sec:Pc_OMP}),
as illustrated in Fig.~\ref{fig:main_path}(b).

\item Because of the non-monotonicity of $c$,
the path from the origin to $\signal{\beta}^*$
is separated into three $\Pct{0.7}$ paths.
One of the separation points is located 
at the breakpoint $[2,0]^{\sf T}$,
and the other one is located at the point where $c$ starts to increase
in Fig.~\ref{fig:main_path}(b).
The separation points of the $\Pct{0.7}$ paths are indicated by squares
      in Figs.~\ref{fig:main_path} and \ref{fig:c_phi}.

\item From the breakpoint $[2,0]^{\sf T}$ to the OLS solution
      $\signal{\beta}^*=[2,1]^{\sf T}$, the local minimality in $\Pc{0.7}$
changes
at the separation point.
All points on the blue curve in Fig.~\ref{fig:main_path}(a)
are local minima in $\Pc{0.7}$, while the red curve
excluding the endpoints contains no local minima in $\Pc{0.7}$.
See the discussion below item 5).

\item Neither the red nor the blue curves ($\Pct{0.7}$ paths)
in Fig.~\ref{fig:main_path}(a)
is composed of $\Qlt{0.7}$ paths.
From $[2,0]^{\sf T}$ to $\signal{\beta}^*=[2,1]^{\sf T}$,
the two $\Qlt{0.7}$ paths are connected at the triangle
where $\beta_1$ starts to increase.
The separation points of $\Qlt{0.7}$ paths are indicated by triangles
      in Figs.~\ref{fig:main_path} and \ref{fig:c_phi}.
The local minimality in $\Ql{0.7}$ changes
at the separation point.
See the discussion below item 5).

\item The $\Pc{0.7}$ global path is discontinuous.
This can be seen by observing that
the minimum value of $\varphi$ in Fig.~\ref{fig:c_phi}(b)
switches from the green curve to the blue one
at the intersection of the two curves.
The $\Pc{0.7}$ global solution therefore jumps
from the green curve to the blue one
      in Fig.~\ref{fig:main_path}(a).

\end{enumerate}
To discuss the local optimality of critical points
 $\widetilde{\signal{\beta}}$
on the curve from the breakpoint $[2,0]^{\sf T}$
to the endpoint $\signal{\beta}^*=[2,1]^{\sf T}$
in Fig.~\ref{fig:main_path}(a),
we will analyze the positive definiteness of the Hessian matrix
$\signal{K}(\widetilde{\signal{\beta}})$ 
with Lemma \ref{lemma:ns}.
The matrix $\signal{K}(\widetilde{\signal{\beta}})$ is indeed
not positive semidefinite from the breakpoint up to 
the separation point (triangle) of the $\Qlt{0.7}$ paths and is
positive semidefinite from the separation point to the endpoint,
and thus, item 4) above applies.
However, between the two separation points (the square and the triangle
 on the curve),
$\signal{K}(\widetilde{\signal{\beta}})$ is positive definite
for tangent vectors, leading to item 3) above.
From Fig.~\ref{fig:main_path}(b), it is apparent that
there are two critical points, off the $\beta_1$-coordinate,
corresponding to some $c$ value.
Indeed, there is another critical point, on the $\beta_1$-coordinate,
corresponding to such a $c$ value.
This implies that, given a surface $\partial \mathcal{B}_c$ for some $c$,
there exist three contours $\mathcal{R}$ of $\varphi$, touching $\partial \mathcal{B}_c$.
In particular, one of the contours $\mathcal{R}$,
passing through a critical point $\widetilde{\signal{\beta}}$ (on the red curve)
very close to the $\beta_1$-coordinate,
is closer to the tangent line than $\partial \mathcal{B}_c$ in the
vicinity of $\widetilde{\signal{\beta}}$, meaning that
$\widetilde{\signal{\beta}}$ is a local maximum in $\Pc{0.7}$.
\end{example}

\begin{figure}[t]
\centering
\begin{psfrags}
\psfrag{lambda0}[c][1.0]{$\lambda=0.2$}
\psfrag{lambda1}[c][1.0]{$\lambda=0.3$}
\psfrag{lambda2}[c][1.0]{$\lambda=0.385$}
\psfrag{lambda3}[c][1.0]{$\lambda=0.5$}
\psfrag{b}[c][1.0]{$\beta$}
\psfrag{f(beta)}[c][1.0]{$f_{\lambda}(\beta)$}
\begin{center}
 \includegraphics[width=8cm]{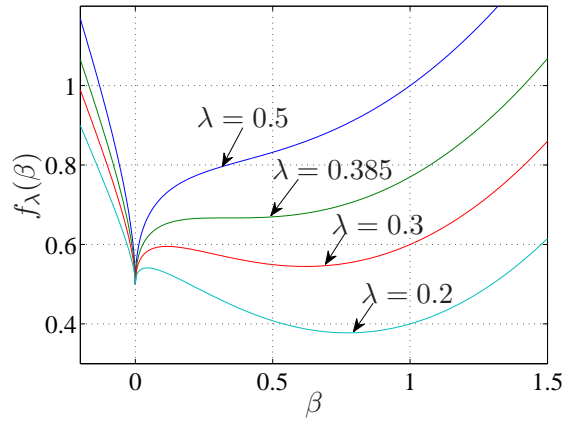} 
\end{center}
\end{psfrags}
\caption{Graphs of $f_{\lambda}(\beta):=\frac{1}{2}(\beta-1)^2 + 2\lambda
 |\beta|^{0.5}$.}
\label{fig:discontinuity}
\end{figure}

\begin{figure}[t]
\centering
\begin{psfrags}
\subfigure[]{
\psfrag{b}[c][1.0]{$\beta$}
\psfrag{b1}[c][1.0]{$\beta_1$}
\psfrag{b2}[c][1.0]{$\beta_2$}
\psfrag{b3}[c][1.0]{$\beta_3$}
\psfrag{b4}[c][1.0]{$\beta_4$}
\psfrag{b5}[c][1.0]{$\beta_5$}
\psfrag{l}[c][1.0]{$\lambda$}
 \includegraphics[width=8cm]{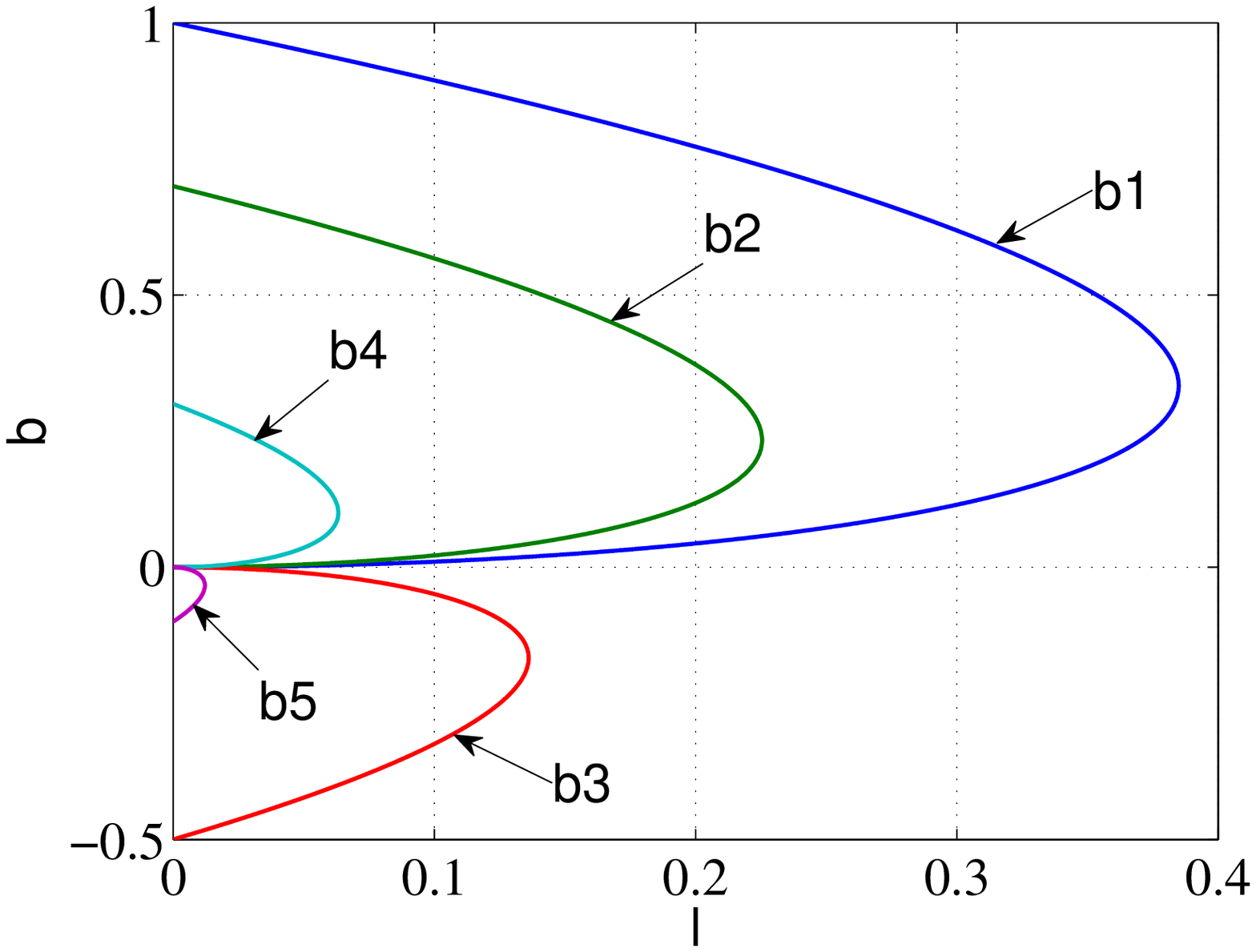}
}
\subfigure[]{
\psfrag{b}[c][1.0]{$\beta$}
\psfrag{b1}[c][1.0]{$\beta_1$}
\psfrag{b2}[c][1.0]{$\beta_2$}
\psfrag{b3}[c][1.0]{$\beta_3$}
\psfrag{b4}[c][1.0]{$\beta_4$}
\psfrag{b5}[c][1.0]{$\beta_5$}
\psfrag{c}[c][1.0]{$c$}
 \includegraphics[width=8cm]{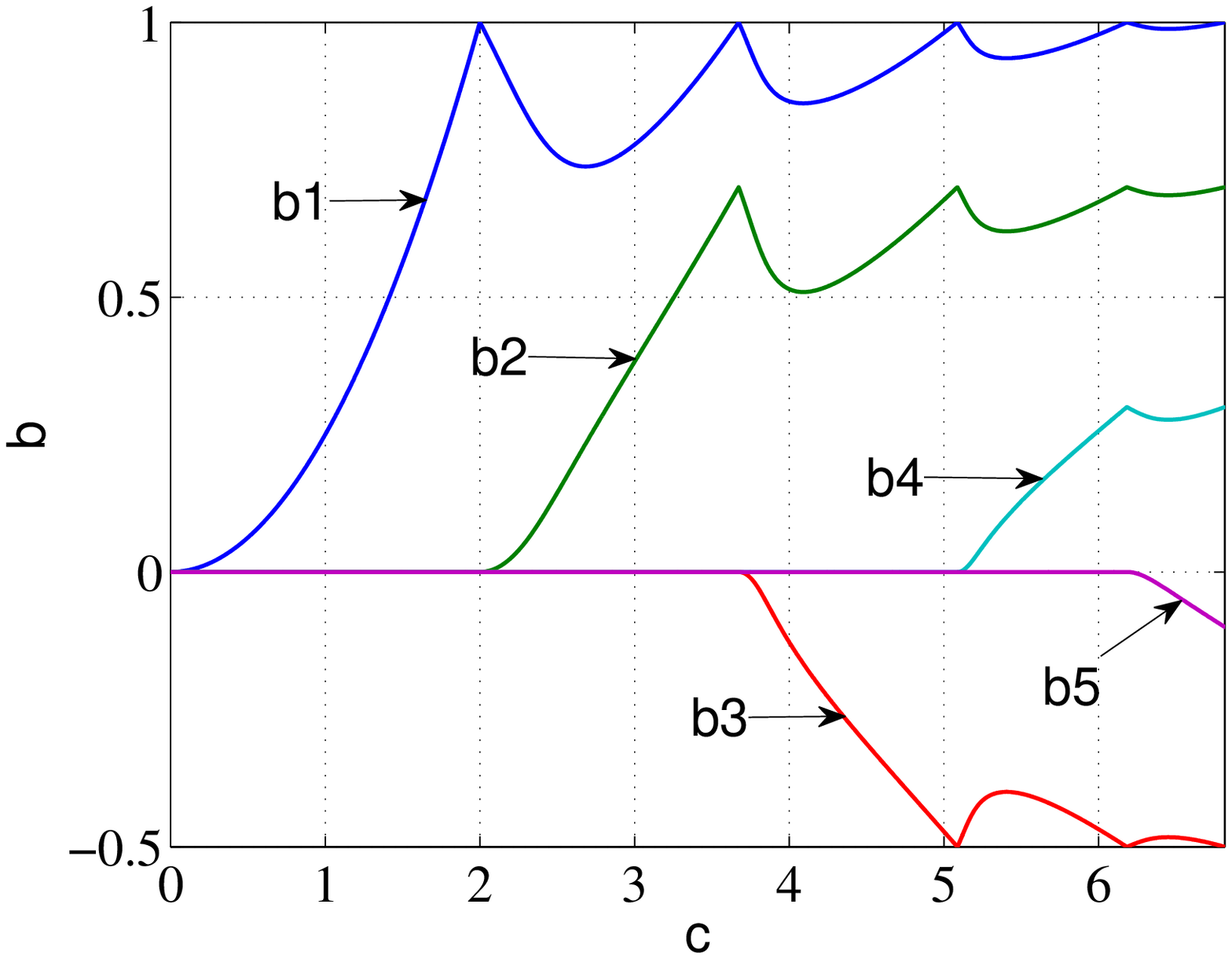}
}
\end{psfrags}
\caption{(a) $\lambda$ - $\beta$ correspondences and (b) the greedy path
 for $n=5$ in Example \ref{example:5D}.}
\label{fig:beta_c}
\end{figure}

\begin{example}[3D non-orthogonal case]
\label{example:counter_example_3D}
Consider the following three-dimensional case:
$\signal{\beta}^*:=[0.2, 0.8, 1]^{\sf T}$
$\signal{G}:=\left[\begin{array}{ccc}
1		    & -0.7 & -0.6 \\
-0.7		    & 1&-0.1 \\
-0.6		    & -0.1 &1 \\
			 \end{array}\right]$, and $p=0.5$.\footnote{
This is the case in which LARS requires the Lasso modification to 
obtain the Lasso solution path.}
In this case, 
$\signal{\nabla}\varphi(\signal{0})=[0.96, -0.56, -0.8]^{\sf T}$
and, hence, \refeq{eq:path_characterization} suggest
the direction $[-\partial_1\varphi(\signal{0}),0,0]^{\sf T}
\propto [-1,0,0]^{\sf T}$ although $\beta_1^*=0.2>0$.
The (unmodified) greedy path is therefore located on the opposite side of the
$\beta_2$-$\beta_3$ coordinate plane from $\signal{\beta}^*$,
and thus the modified greedy path selects another direction 
$[0,0,-\partial_3\varphi(\signal{0})]^{\sf T}
\propto [0,0,1]^{\sf T}$.
The modified greedy path leads to the OLS solution
$\signal{\beta}^*$ via the breakpoints
$\signal{\beta}_{\rm BR}^1=[0,0,0.8]^{\sf T}$ and
$\signal{\beta}_{\rm BR}^2\approx [0,0.6465,0.8646]^{\sf T}$.
This is actually the main path.
The unmodified greedy path passes through the breakpoints
 $\check{\signal{\beta}}_{\rm BR}^{1}:=[-0.96,0,0]^{\sf T}$ and
 $\check{\signal{\beta}}_{\rm BR}^{2}:=[-0.75,0,0.35]^{\sf T}$,
and then all the components become active,
ending up with two active components $\beta_1$ and $\beta_2$
simultaneously going back to zero at $[0,0,0.8]^{\sf T}$.
\end{example}
\clearpage
\begin{example}[5D orthogonal case]
\label{example:5D}
Consider the following case of dimension $n=5$
under the orthogonality condition
$\signal{X}\signal{X}^{\sf T}=\signal{I}$:
 $\varphi(\signal{\beta}):=\frac{1}{2}
\norm{\signal{\beta} - \signal{\beta}^*}_2^2$ with
$\signal{\beta}^*:=[1, 0.7, -0.5, 0.3, -0.1]^{\sf T}$
and $p=0.5$.
The function $f_{\lambda}$ can be separated as follows:
\begin{equation}
f_{\lambda}(\signal{\beta})=
\sum_{i=1}^{n} f_{\lambda,i}(\beta_i), \ \signal{\beta}\in\real^n,
\label{eq:f_lambda_separate2}
\end{equation}
where
$f_{\lambda,i}(\beta):= \frac{1}{2}(\beta - \beta_i^*)^2 +
 \frac{1}{p}\lambda\abs{\beta}^{p}$,
$\beta\in\real$.
The critical-point condition for $\Ql{p}$ can also be
written separately as follows (see Definition \ref{def:critical_point}):
\begin{equation}
 \signal{\nabla}_{\mathcal{I}}f_{\lambda}(\signal{\beta})
=\left[
\begin{array}{c}
f'_{\lambda,i_1}(\beta_{i_1}) \\
f'_{\lambda,i_2}(\beta_{i_2}) \\
\vdots \\
f'_{\lambda,i_s}(\beta_{i_s}) \\
\end{array}
\right]=\signal{0}_{i_s},
\end{equation}
where $\mathcal{I}:=\{i_1,i_2,\cdots,i_s\}:={\rm supp}(\signal{\beta})$,
$f'_{\lambda,i_s}:=\frac{d}{d\beta}f_{\lambda,i_s}$, and
$\signal{0}_{i_s}$ is the zero vector of length $i_s$.

The nonzero critical points for each individual function $f_{\lambda, i_s}$
are plotted in Fig.~\ref{fig:beta_c}(a).
(Note that zero is always a critical point for any $f_{\lambda,i_s}$
and thus is omitted.)
On each curve in Fig.~\ref{fig:beta_c}(a),
there are two points corresponding to each $\lambda$.
The one with a smaller absolute value is a local maximum and
the one with a larger absolute value is a local minimum
(see Fig.~\ref{fig:discontinuity}).
The greedy path goes along the $\beta_1$ coordinate until
$\beta_1=\beta_1^*=1$.
In Fig.~\ref{fig:beta_c}(a), we can trace the blue curve from $(0,0)$ to
$(0,1)$,
and 
in Fig.~\ref{fig:beta_c}(b) we can trace the blue curve from $(0,0)$ to
$(2,1)$.
(The variables $\beta_2$, $\beta_3$, $\beta_4$, and $\beta_5$ stay at
zero.)
Next, the new entry $\beta_2$ becomes active, and it increases from zero
up to $\beta_2=\beta_2^*=0.7$.
In this case, we can trace the blue curve in Fig.~\ref{fig:beta_c}(b) 
from (0,1) to the next peak and
the green curve from (2.0) to its first peak.
In Fig.~\ref{fig:beta_c}(a), we can trace the blue curve from $(0,1)$
until $\lambda$ reaches a point at which the function $f_{\lambda,2}$
has its unique nonzero critical point (the peak of the green curve)
and trace the same path in a reverse way back to $(0,1)$.
Also in Fig.~\ref{fig:beta_c}(a), we can trace the green curve from $(0,0)$
to $(0,0.7)$.
(The variables $\beta_3$, $\beta_4$, and $\beta_5$ stay at
zero meanwhile.)
One can follow the same procedure to see the whole picture of the greedy path.
It can be seen that the greedy path connects the origin and the OLS solution
 $\signal{\beta}^*$ continuously in this case.

All the critical paths can be found in this way.
For instance,
let us consider another particular path on which
all the variables become active when stepping slightly away from the
origin.
In this case, $\beta_5$ achieves its peak in Fig.~\ref{fig:beta_c}(a) 
before the others
and one cannot increase $\lambda$ any
further.
What one can do here is to reduce $\lambda$.
Accordingly, $\beta_1$, $\beta_2$, $\beta_3$, and $\beta_4$ can only go back
to zero by tracing the same path in Fig.~\ref{fig:beta_c}(a) in a reverse way,
and only $\beta_5$ can trace the purple curve up to $(0,-0.1)$.
In this case, the whole path starts at the origin and ends at
$\check{\signal{\beta}}:=[0,0,0,0,-0.1]^{\sf T}$;
it consists of two $\Qlt{0.5}$ paths because 
a critical path is a single valued function of $c$ (or $\lambda$) by
definition.
Along the path, $c$ increases up to some point and then starts
to decrease.
Hence, the path is divided into two parts: the part containing the origin
is a $\Pct{0.5}$ path; the other part becomes another $\Pct{0.5}$
path by extending it with a straight line to the origin along the $\beta_5$ coordinate.
\end{example}





\bibliographystyle{IEEEtran}
\bibliography{Lp}




%








\clearpage

\end{document}